\documentclass[aps,prd,amsmath,floats,floatfix,twocolumn,superscriptaddress,nofootinbib,showpacs]{revtex4}

\usepackage{amssymb}
\usepackage{amsmath}
\usepackage{verbatim}
\usepackage{mathrsfs}
\usepackage{amsfonts}
\usepackage{latexsym}
\usepackage{epsfig}
\usepackage{color}
\usepackage{graphicx,subfigure}
\usepackage{units}
\usepackage{overpic}

\begin{document}


\definecolor{orange}{rgb}{0.9,0.45,0} 

\newcommand{\argelia}[1]{\textcolor{red}{{\bf Argelia: #1}}}
\newcommand{\dario}[1]{\textcolor{red}{{\bf Dario: #1}}}
\newcommand{\juanc}[1]{\textcolor{green}{{\bf JC: #1}}}
\newcommand{\juan}[1]{\textcolor{cyan}{{\bf Juan B: #1}}}
\newcommand{\alberto}[1]{\textcolor{blue}{{\bf Alberto: #1}}}
\newcommand{\miguela}[1]{\textcolor{red}{{\bf MiguelA: #1}}}
\newcommand{\mm}[1]{\textcolor{orange}{{\bf MM: #1}}}
\newcommand{\OS}[1]{\textcolor{blue}{{\bf Olivier: #1}}}

\long\def\symbolfootnote[#1]#2{\begingroup%
\def\thefootnote{\fnsymbol{footnote}}\footnote[#1]{#2}\endgroup}


\title{Dynamical evolutions of $\ell$-boson stars in spherical symmetry}

\author{Miguel Alcubierre}
\affiliation{Instituto de Ciencias Nucleares, Universidad Nacional Aut\'onoma de M\'exico,
Circuito Exterior C.U., A.P. 70-543, M\'exico D.F. 04510, M\'exico}

\author{Juan Barranco}
\affiliation{Departamento de F\'isica, Divisi\'on de Ciencias e Ingenier\'ias,
Campus Le\'on, Universidad de Guanajuato, Le\'on 37150, M\'exico}

\author{Argelia Bernal}
\affiliation{Departamento de F\'isica, Divisi\'on de Ciencias e Ingenier\'ias,
Campus Le\'on, Universidad de Guanajuato, Le\'on 37150, M\'exico}

\author{Juan Carlos Degollado}
\affiliation{Instituto de Ciencias F\'isicas, Universidad Nacional Aut\'onoma de M\'exico,
Apdo. Postal 48-3, 62251, Cuernavaca, Morelos, M\'exico}

\author{Alberto Diez-Tejedor}
\affiliation{Departamento de F\'isica, Divisi\'on de Ciencias e Ingenier\'ias,
Campus Le\'on, Universidad de Guanajuato, Le\'on 37150, M\'exico}

\author{Miguel Megevand}
\affiliation{Instituto de F\'isica Enrique Gaviola, CONICET. Ciudad Universitaria, 5000 C\'ordoba, Argentina}

\author{Dar\'io N\'u\~nez}
\affiliation{Instituto de Ciencias Nucleares, Universidad Nacional Aut\'onoma de M\'exico,
Circuito Exterior C.U., A.P. 70-543, M\'exico D.F. 04510, M\'exico}

\author{Olivier Sarbach}
\affiliation{Instituto de F\'isica y Matem\'aticas, Universidad Michoacana de San Nicol\'as de Hidalgo,
Edificio C-3, Ciudad Universitaria, 58040 Morelia, Michoac\'an, M\'exico}


\date{\today}


\begin{abstract}
In previous work, we have found new static, spherically symmetric
boson star solutions which generalize the standard boson stars by
allowing a particular superposition of scalar fields in which each of
the fields is characterized by a fixed value of its non-vanishing
angular momentum number $\ell$. We call such solutions ``$\ell$-boson
stars''. Here, we perform a series of fully non-linear dynamical
simulations of perturbed $\ell$-boson stars in order to study their
stability, and the final fate of unstable configurations. We show that
for each value of $\ell$, the configuration of maximum mass separates
the parameter space into stable and unstable regions.  Stable
configurations, when perturbed, oscillate around the unperturbed
solution and very slowly return to a stationary
configuration. Unstable configurations, in contrast, can have three
different final states: collapse to a black hole, migration to the
stable branch, or explosion (dissipation) to infinity. Just as it
happens with $\ell=0$ boson stars, migration to the stable branch or
dissipation to infinity depends on the sign of the total binding
energy of the star: bound unstable stars collapse to black holes or
migrate to the stable branch, whereas unbound unstable stars either
collapse to a black hole or explode to infinity.  Thus, the parameter
$\ell$ allows us to construct a new set of stable configurations.  All
our simulations are performed in spherical symmetry, leaving a more
detailed stability analysis including non-spherical perturbations for
future work.
\end{abstract}




\maketitle


\section{Introduction}
\label{sec:intro}

Scalar fields are an ubiquitous theme in modern
cosmology~\cite{Matos:2000ss,Schive:2014dra,Hui:2016ltb,Liddle2000,Caldwell:1997ii}.
They can also form localized, globally regular, self-gravitating
configurations.  Depending on the scalar field properties, different
types of configurations have been found such as
oscillatons~\cite{Seidel91} (for real scalar fields), boson stars
(BSs)~\cite{Kaup68, Ruffini69}, BSs with
self-interaction~\cite{Colpi86}, multistate
BSs~\cite{Bernal:2009zy}, and multi-oscillating BSs~\cite{Choptuik:2019zji}.
Further studies have shown that BS's may
also have electric charge and/or rotation.  Since then many works have
been dedicated to study their nature and their potential detection as
BHs mimickers (see~\cite{Schunck:2003kk,Liebling:2012fv} for reviews).

For some values of the mass of the constituent boson, BSs are
indistinguishable from BHs in the weak field region. For this reason,
BSs have been considered as alternatives for the central galactic
BHs~\cite{Torres2000,Guzman:2009zz,AmaroSeoane:2010qx}.  It is then
important to be able to differentiate between the two objects, which can
both account for current observable constraints. 
Despite the fact that recent observations of the center of the galaxy M87 by the Event Horizon Telescope~\cite{Akiyama:2019cqa} have already put important constraints on black hole alternatives (including BSs~\cite{Olivares:2018abq,Akiyama:2019fyp}), there is still a need to better understand their physical properties to accurately model their associated accretion flow and resulting image.

Recently we have found new BS solutions that we have called
$\ell$-boson stars~\cite{Alcubierre:2018ahf}.\footnote{We have
  recently become aware of related work by Mielke and Scherzer in
  1981~\cite{Mielke81}, where similar configurations of the
  Einstein-Klein-Gordon system are studied. In their work, they also
  consider families of complex scalar fields. However, their procedure
  involves an averaging of the stress-energy tensor over the sphere in
  order to get rid of the angular dependency, whereas in our work the
  family is deliberately chosen such that the angular dependency is
  exactly cancelled in the total stress-energy tensor without taking
  averages. We thank E. Mielke for pointing Ref.~\cite{Mielke81} to
  us.} These $\ell$-boson stars are compact spherically symmetric
configurations composed by an odd number of complex scalar fields. The
configurations are parametrized by an angular momentum number $\ell$,
hence the name, with the $\ell=0$ case corresponding to the standard
well-known boson stars.  In particular, $\ell$-boson stars can be
extremely compact objects, approaching the Buchdahl limit\footnote{See
  Ref.~\cite{Andreasson:2007ck} for a rather general result regarding
  the Buchdahl limit and generalizations thereof for static,
  spherically symmetric spacetimes. Interestingly, the $\ell$-boson
  stars discussed in this article {\em do not} satisfy the assumption
  of a non-negative radial pressure made in that reference.}  $M/R<4/9
\sim 0.44$ (for $\ell=4$ we have found solutions with $M/R \sim 0.3$).

Any astrophysical model for a compact object, in order to be viable,
must be stable or at least stable for a sufficiently long time. In
this work we therefore focus on studying the stability properties of
$\ell$-boson stars.  In order to test the nonlinear (in)stability of
these solutions we perform fully nonlinear numerical simulations in
spherical symmetry.  We use as initial data the configurations found
in~\cite{Alcubierre:2018ahf}, and perturb them in several ways, namely
adding or subtracting a small quantity of mass while trying to keep
the total number of particles fixed. We have found that stable
configurations, when perturbed, oscillate around the unperturbed
solution and seem to very slowly return to a stationary
configuration. Unstable configurations, in contrast, can have three
different final states: collapse to a black hole, migration to the
stable branch, or explosion (dissipation) to infinity.  Just as it
happens with $\ell=0$ boson stars, migration to the stable branch or
dissipation to infinity depend on the sign of the total binding energy
of the star: bound unstable stars collapse to black holes or migrate
to the stable branch, whereas unbound unstable stars either collapse
to a black hole or explode to infinity. It is important to mention
that for both stable configurations, and unstable configurations that
migrate to the stable branch, the relaxation times seem to be
extremely long.  In some cases we have followed these configurations
for thousands of light crossing times (millions of numerical time
steps) and, although we can be quite confident that the final state is
indeed stable, at this point we can not rule out the possibility that
those configurations will not settle down to a single frequency
$\ell$-boson star, but will settle instead to some form of
multi-oscillating solution such as those recently studied
in~\cite{Choptuik:2019zji}.

The paper is organized as follows: In Section \ref{sec:EinsteinKG} we
describe the Einstein Klein-Gordon system and the decomposition of the
fields required to have a spherically symmetric configuration.  In
Section~\ref{sec:stationary} we review the stationary solutions found
in~\cite{Alcubierre:2018ahf} that are used as initial data in our
numerical evolutions. In Section~\ref{sec:perturbations} we give a
description of the type of perturbations we apply to the static
solutions. In Section~\ref{sec:numerical} we present the analysis
tools and numerical techniques used in our simulations. We present our
results in section~\ref{sec:results}. Finally, in
Section~\ref{sec:conclusions} we give some concluding remarks.


\section{The Einstein Klein-Gordon system}
\label{sec:EinsteinKG}

We will consider a collection of an arbitrary odd number of complex,
non-interacting scalar fields of equal mass $\mu$, minimally coupled
to gravity. Following~\cite{Olabarrieta:2007di,Alcubierre:2018ahf} we
consider solutions of the form
\begin{equation}
\Phi_{\ell m}(t,r,\vartheta,\varphi) = \phi_\ell(t,r) Y^{\ell m}(\vartheta,\varphi),
\label{eq:Ylm}
\end{equation}
where the angular momentum number $\ell$ is fixed, and $m$ takes
values $m = -\ell,-\ell+1,\ldots,\ell$. As usual $Y^{\ell m}$ denotes
the standard spherical harmonics, and the amplitudes $\phi_\ell(t,r)$
are the {\it same} for all $m$. As already shown
in~\cite{Olabarrieta:2007di,Alcubierre:2018ahf}, this leads to a total
stress energy-momentum tensor which is spherically symmetric.

In order to solve the field equations, we will then consider
a spherically symmetric spacetime with a line element given by:
\begin{equation}
ds^2 = -\alpha^2 dt^2 + \psi^4 \left( A dr^2
+ r^2 B d\Omega^2 \right) \; ,
\label{eq:metric}
\end{equation}
where $(\alpha,A,B,\psi)$ are functions of $(r,t)$ only, and
$d\Omega^2$ is the standard solid angle element. Notice that this form
of the metric might seem too generic, and in order to find boson star
initial data one typically takes $\psi=B=1$.~\footnote{Note that the
  form of the metric used here differs from the one of
  Ref.~\cite{Alcubierre:2018ahf}. In particular, in that reference we
  used $\psi=B=1$, and $\gamma=\sqrt{A}$.}  However, this is the form
of the metric we will use for our dynamical simulations below, since
we will be using a spherically symmetric version of the
Baumgarte-Shapiro-Shibata-Nakamura (BSSN)
formulation~\cite{Baumgarte:1998te,Shibata95,Brown:2009dd,Alcubierre:2010is}.
With these assumptions and definitions, the Klein-Gordon equation can
be written in first order form as (for simplicity we will suppress the
index $\ell$):
\begin{eqnarray}
\partial_t \phi &=& \alpha \Pi \; , \\
\partial_t \chi &=& \alpha \partial_r \Pi + \Pi \partial_r \alpha \; , \\
\partial_t \Pi  &=& \frac{\alpha}{A \psi^4} \left[ \partial_r \chi
+ \chi \left( \frac{2}{r} - \frac{\partial_r A}{2A}
+ \frac{\partial_r B}{B} + 2 \: \partial_r \psi \right) \right] \nonumber \\
&+& \frac{\chi \: \partial_r \alpha}{A \psi^4} + \alpha K \Pi
- \alpha \left( \mu^2 + \frac{\ell (\ell+1)}{r^2B \psi^4} \right) \; ,
\end{eqnarray}
with $K=K^m_m$ the trace of the extrinsic curvature of the
hypersurfaces of constant time, and where we have
defined:~\footnote{Again, note that the definition of $\chi$ in
    Eq.~(\ref{Eq:chiPi}) differs from the definition of the quantity
    $\chi_\ell$ in~\cite{Alcubierre:2018ahf} by a factor of
    $\gamma=\sqrt{A}$.}
\begin{equation}
\chi := \partial_r \phi \; , \qquad
\Pi := \frac{\partial_t \phi}{\alpha} \; .
\label{Eq:chiPi}
\end{equation}

Furthermore, the stress-energy tensor can be shown to take the
form~\cite{Alcubierre:2018ahf}:
\begin{eqnarray}
T_{mn} &=& (\partial_m\phi)^*(\partial_n\phi) 
 - \frac{\gamma_{mn}}{2} \left[ \rule{0mm}{5mm} \gamma^{pq}(\partial_p\phi)^*(\partial_q\phi) \right. \nonumber \\
 &+& \left. \left( \mu^2 + \frac{\ell(\ell+1)}{r^2} \right) |\phi|^2 \right] \; , \nonumber \\
T_{mN} &=& 0 \; ,\\
T_{MN} &=& - r^2 B \psi^4 \frac{\gamma_{MN}}{2}
\left[ \gamma^{pq}\left(\partial_p\phi\right)^*\left(\partial_q\phi\right)
+ \mu^2 |\phi|^2 \right] \; ,\nonumber
\end{eqnarray}
where here $(m,n) = (t,r)$, $(M,N) = (\vartheta,\varphi)$,
$\gamma_{mn}$ is the 2D time--radial metric, \mbox{$\gamma_{mn} dx^m
  dx^n = -\alpha^2 dt^2 + A \psi^4 dr^2$}, and $\gamma_{MN}$ the 2D
angular metric for the unit sphere, \mbox{$\gamma_{MN} dx^M dx^N =
  d\Omega^2$}.

Notice that the normalization in the above expressions for the
stress-energy tensor differs from the one used in
reference~\cite{Alcubierre:2018ahf} by a factor of $(2\ell
+1)/4\pi$. The reason for this is that we have absorbed that factor
into the definition of $\phi$ in order to be consistent with the
normalization used in the numerical evolution code, which takes $\phi$
as a single scalar field with a modified evolution equation and
stress-energy tensor, instead of a sum over $(2\ell+1)$
 independent fields (the extra factor of $4\pi$ comes from
the normalization of the spherical harmonics).  With the normalization
above the Einstein field equations take the standard form $G_{\mu \nu}
= 8 \pi T_{\mu \nu}$ (we use Planck units such that $G=c=\hbar=1$).

With the expressions above for the stress-energy tensor, the energy
density $\rho_E$, momentum density $P_i$ and stress tensor $S_{ij}$ as
seen by the normal (Eulerian) observers become:
\begin{eqnarray}
\rho_E &=& n^\mu n^\nu T_{\mu \nu} \nonumber \\
&=& \frac{1}{2} \left[ |\Pi|^2 + \frac{|\chi|^2}{A \psi^4} 
+ \left( \mu^2 + \frac{\ell(\ell+1)}{r^2} \right) |\phi|^2 \right] , \label{eq:energydensity} \\
P_r &=& - n^\mu T_{r \mu} = - \frac{1}{2} \left( \chi \Pi^* + \Pi \chi^* \right) , \label{eq:momentumdensity} \\
S^r_r &=& \frac{1}{2} \left[ |\Pi|^2 + \frac{|\chi|^2}{A \psi^4} 
- \left( \mu^2 + \frac{\ell(\ell+1)}{r^2} \right) |\phi|^2 \right] , \hspace{5mm}  \\
S^\theta_\theta &=& \frac{1}{2} \left[ |\Pi|^2 - \frac{|\chi|^2}{A \psi^4}
- \mu^2 |\phi|^2 \rule{0mm}{5mm} \right] ,
\end{eqnarray}
where $n^\mu=(1/\alpha,0,0,0)$ is the unit normal vector to the
spatial hypersurfaces. In particular, the momentum density is purely
radial because of the spherical symmetry.


\section{Stationary initial data}
\label{sec:stationary}

We will consider stationary $\ell$-boson stars, with a complex scalar
field that has the form:
\begin{equation}
\phi(r,t) = \varphi(r) e^{i \omega t} \; ,
\label{eq:ansatz}
\end{equation}
with $\omega$ and $\varphi(r)$ real-valued. At $t=0$ this implies:
\begin{eqnarray}
\phi(r,t=0) &=& \varphi \;, \\
\chi(r,t=0) &=& \varphi' \; , \\
\Pi(r,t=0)  &=& i \omega \varphi / \alpha \; .
\end{eqnarray}
We then see that the initial scalar field and its radial derivative
are purely real, while the initial time derivative is purely
imaginary.

The initial data for stationary $\ell$-boson stars was discussed in
detail in~\cite{Alcubierre:2018ahf}. As discussed there, even though
the scalar field oscillates in time the stress-energy tensor is
time-independent, and these objects result in static solutions to the
Einstein field equations.  In order to find initial data, one
substitutes the ansatz~\eqref{eq:ansatz} in the Klein-Gordon equation,
assumes that the spatial metric is in the areal gauge so that
$\psi=B=1$ in the metric~\eqref{eq:metric} above, asks for the
spacetime metric to be static so that the extrinsic curvature $K_{ij}$
vanishes, and solves the Hamiltonian constraint for the radial metric
$A$.  For the lapse function $\alpha$ we use the ``polar slicing''
condition, which asks for the time derivative of the angular component
of the extrinsic curvature to vanish, $\partial_t K_{\theta
  \theta}=0$, and results in a first order differential equation for
the lapse function $\alpha$. This results in the following system of
three equations (Klein-Gordon equation, Hamiltonian constraint and
polar slicing condition respectively):
\begin{eqnarray}
\partial^2_r \varphi &=& -  \partial_r \varphi \left( \frac{2}{r} + \frac{\partial_r \alpha}{\alpha}
- \frac{\partial_r A}{2A} \right)  \nonumber \\
&+& A \varphi \left( \mu^2 + \frac{\ell(\ell+1)}{r^2} - \frac{\omega^2}{\alpha^2} \right) \: , \\
\partial_r A &=& A \left\{ \frac{(1-A)}{r} + 4 \pi r A \left[ \frac{(\varphi')^2}{A} \right. \right. \nonumber \\
&+& \left. \left. \varphi^2 \left( \mu^2 + \frac{\ell(\ell+1)}{r^2}
+ \frac{\omega^2}{\alpha^2} \right) \right] \right\} \: , \\
\partial_r \alpha &=& \alpha \left[ \frac{(A-1)}{r} + \frac{\partial_r A}{2A} \right. \nonumber \\
&-& \left. 4 \pi r A \varphi^2 \left( \mu^2 + \frac{\ell(\ell+1)}{r^2} \right) \right] \: .
\end{eqnarray}
Notice that instead of polar slicing one could ask for a maximal
slicing condition, $\partial_t K=0$, which in this case can be shown
to be equivalent, but results in a second order differential
equation for $\alpha$ instead.

By analyzing the Klein-Gordon equation one finds that for small $r$
the scalar field behaves as \mbox{$\varphi \sim \varphi_0 r^\ell$}.
For a fixed value of $\ell$, and a given value of the parameter
$\varphi_0$, the above system of equations becomes a nonlinear eigenvalue
problem for the frequency $\omega$, subject to the boundary condition
that $\varphi$ decays exponentially far away.

In~\cite{Alcubierre:2018ahf} it was also found that $\ell$-boson stars
possess similar properties to those of standard $\ell=0$ boson
stars. Specifically, for a given angular momentum number $\ell$, as
the parameter $\varphi_0$ increases, the equilibrium configurations
exhibit a maximum value of the mass, and this maximum grows with
$\ell$, leading to more compact objects. Also, for each value of
$\ell$, the space of solutions separates in two distinct branches to
either side of the maximum mass.  We will show below that, just as occurs
with the $\ell=0$ case, those two branches correspond to stable
and unstable configurations.

Before introducing perturbations to the initial data for the
$\ell$-boson stars, it is important to consider three physical
quantities related to the scalar field that are important in
characterizing the different solutions. The first two are the energy
density $\rho_E$ (given by equation~\eqref{eq:energydensity} above),
and the boson (particle) density $\rho_B$:
\begin{eqnarray}
\rho_E &=& \frac{1}{2} \left[ |\Pi|^2 + \frac{|\chi|^2}{A \psi^4} 
+ \left( \mu^2 + \frac{\ell(\ell+1)}{r^2} \right) |\phi|^2 \right] ,
\hspace{5mm} \label{eq:Edensity} \\
\rho_B &=& - n^\mu J_\mu = \phi_R \Pi_I - \phi_I \Pi_R \: ,
\label{eq:Bdensity}
\end{eqnarray}
where $J^\mu$ is the conserved particle current
\begin{equation}
J^\mu = - \frac{1}{2} \: \operatorname{Im} \left( \phi^* \nabla^\mu \phi
- \phi \: \nabla^\mu \phi^* \right) \: ,
\label{eq:Jcurrent}
\end{equation}
and where
the sub-indices $R$ and $I$ refer to the real and imaginary parts
respectively. Substituting the ansatz~\eqref{eq:ansatz} these
expressions reduce to:
\begin{eqnarray}
\rho_E &=& \frac{1}{2} \left[ \frac{(\varphi')^2}{A \psi^4}
+ \left( \mu^2 + \frac{\ell(\ell+1)}{r^2} 
+ \frac{\omega^2}{\alpha^2} \right) \varphi^2 \right] , \hspace{5mm}
\label{eq:Edensity0} \\
\rho_B &=& \omega \varphi^2 / \alpha \: .
\label{eq:Bdensity0}
\end{eqnarray}
Notice that both these quantities are clearly time independent.

The third quantity we need to consider is the momentum density of the
scalar field, which is given by equation~\eqref{eq:momentumdensity}
above, and has the form:
\begin{equation}
P_r = - \frac{1}{2} \left( \chi \Pi^* + \Pi \chi^* \right)
= - \chi_R \Pi_R - \chi_I \Pi_I \; .
\label{eq:Mdensity}
\end{equation}
This can now be easily shown to vanish when we substitute the
ansatz~\eqref{eq:ansatz}. Notice that, since the spacetime metric for
the $\ell$-boson stars is static and the momentum density vanishes,
the momentum constraint is trivially satisfied and can be safely
ignored when solving for the initial data.


\section{Perturbed initial data}
\label{sec:perturbations}

We will now add to the stationary initial data described above small
(but non-linear) perturbations, such that at $t=0$ we will have:
\begin{eqnarray}
\phi_R = \varphi_0 + \delta \varphi_R \; , &\qquad& \phi_I = \delta \varphi_I \; , \\
\Pi_R = \delta \Pi_R \; , \hspace{6mm} &\qquad& \Pi_I = (\Pi_I)_0 + \delta \Pi_I \; ,
\end{eqnarray}
where the subindex $0$ refers to the unperturbed solution. Notice
that in particular we have $(\Pi_I)_0= \omega \varphi_0/ \alpha_0$, with
$\alpha_0$ the unperturbed lapse and $\omega$ the frequency of the
unperturbed solution.

We will consider first the effect that this perturbation has on the
momentum density.  The reason for this is that the initial data, once
perturbed, will not correspond any more to a static spacetime.
However, for simplicity, we would like to ask for the initial data to
be time symmetric, so that we can still ignore the momentum constraint
and only solve the Hamiltonian constraint at $t=0$.  But in order to
do this we must ask for the perturbation to keep the initial momentum
density of the scalar field equal to zero.

Substituting the perturbation in the expression for the momentum
density~\eqref{eq:Mdensity}, and remembering that \mbox{$\chi_{I,R} :=
  \partial_r \varphi_{I,R}$}, we find:
\begin{equation}
P_r = - \varphi_0' \delta \Pi_R - (\Pi_I)_0 \delta \chi_I
- \delta \Pi_R \delta \chi_R - \delta \Pi_I \delta \chi_I \; .
\end{equation}  In order for $P_r$ to vanish we must then
ask for:
\begin{equation}
\varphi_0' \delta \Pi_R + (\Pi_I)_0 \delta \chi_I
+ \delta \Pi_R \delta \chi_R + \delta \Pi_I \delta \chi_I = 0 \; .
\end{equation}
If we assume that the perturbations are small, we can ask for the
linear and quadratic terms to vanish separately:
\begin{eqnarray}
\varphi_0' \delta \Pi_R + (\Pi_I)_0 \: \delta \chi_I &=& 0 \; , \\
\delta \Pi_R \delta \chi_R + \delta \Pi_I \delta \chi_I &=& 0 \; .
\end{eqnarray}
In principle there are may ways to satisfy these two conditions.  The
simplest choice is to ask for $\delta \Pi_R = \delta \varphi_I =
\delta \chi_I = 0$, that is, the perturbation must be such that the
initial value of $\phi$ (and hence $\chi$) remains purely real,
while the initial value of $\Pi$ remains purely imaginary.

Consider now the boson density. The perturbed initial data will result in
a boson density given by:
\begin{equation}
\rho_B = (\rho_B)_0 + \varphi_0 \delta \Pi_I + (\Pi_I)_0 \delta \varphi_R
+ \delta \varphi_R \delta \Pi_I - \delta \varphi_I \delta \Pi_R \; ,
\end{equation}
where $(\rho_B)_0$ is the unperturbed boson
density given by~\eqref{eq:Bdensity0}.  If we substitute the condition
\mbox{$\delta \Pi_R = \delta \varphi_I = 0$}, this reduces to:
\begin{equation}
\rho_B = (\rho_B)_0 + \varphi_0 \delta \Pi_I + (\Pi_I)_0 \delta \varphi_R 
+ \delta \varphi_R \delta \Pi_I \; .
\end{equation}

Now, if we want a perturbation such that the boson density remains
unchanged, we need to ask for:
\begin{equation}
\varphi_0 \delta \Pi_I + (\Pi_I)_0 \delta \varphi_R
+ \delta \varphi_R \delta \Pi_I = 0 \; .
\end{equation}
Again, for small perturbations we can ask for the linear and quadratic
terms to vanish separately:
\begin{eqnarray}
\varphi_0 \delta \Pi_I + (\Pi_I)_0 \delta \varphi_R &=& 0 \; , \\
\delta \varphi_R \delta \Pi_I &=& 0 \; .
\end{eqnarray}
We now immediately see that these two conditions can not both be
satisfied at the same time unless both $\delta \varphi_R$ and $\delta
\Pi_I$ vanish, so it is not possible to keep the boson density
constant with these type of perturbations.  However, for small
perturbations we can still keep the linear part equal to zero in two
special cases:

\begin{enumerate}

\item We choose an {\em external}\/ perturbation, that is, one that
  has compact support outside the star, so that both 
  $\phi_0 \delta \Pi_I$ and $(\Pi_I)_0 \delta \varphi_R$ will be
  identically zero.  Notice that physically an external perturbation
  means that we are letting scalar field fall into the boson star from
  the outside.~\footnote{Boson stars in fact do not have a finite
    radius, so one can never have an ``external'' perturbation
    exactly. But the scalar field decays exponentially rapidly so we
    can have a very good approximation to this situation if the
    perturbation is sufficiently far away.}

\item We choose an {\em internal}\/ perturbation to the star such that:
\begin{equation}
\varphi_0 \delta \Pi_I = - (\Pi_I)_0 \delta \varphi_R \; .
\end{equation}
Substituting now the value of $(\Pi_I)_0$ this conditon reduces to
\begin{equation}
\delta \Pi_I = - \frac{\omega}{\alpha_0} \: \delta \varphi_R \; .
\end{equation}

\end{enumerate}

In both these cases, the perturbation will produce only a second order
change in small quantities in the boson density $\delta \rho_B =
\delta \varphi_R \delta \Pi_I$.

Finally, let us consider the effect of the perturbation in the energy
density.  We find:
\begin{eqnarray}
\rho_E &=& (\rho_E)_0 + Q(r) \varphi_0 \delta \varphi_R + (\Pi_I)_0 \delta \Pi_I
+ \frac{\varphi_0' \delta \chi_R}{A_0 \psi_0^4} \nonumber \\
&+& \frac{1}{2} \left[ \delta \Pi_R^2 + \delta \Pi_I^2
+ \frac{1}{A_0 \psi_0^4} \left( \delta \chi_R^2 + \delta \chi_I^2 \right) \rule{0mm}{5mm} \right. \nonumber \\
&+& \left. Q(r)\left( \delta \varphi_R^2 + \delta \varphi_I^2 \right) \rule{0mm}{5mm} \right] \; ,
\end{eqnarray}
with $(\rho_E)_0$ the unperturbed energy density give
by~\eqref{eq:Edensity0}, and where we introduced the shorthand
\mbox{$Q(r):=\mu^2 + \ell(\ell+1)/r^2$}. If we take again $\delta
\Pi_R = \delta \varphi_I = \delta \chi_I = 0$, this reduces to:
\begin{eqnarray}
\rho_E &=& (\rho_E)_0 + Q(r) \varphi_0 \delta \varphi_R + (\Pi_I)_0 \delta \Pi_I
+ \frac{\varphi_0' \delta \chi_R }{A_0 \psi_0^4} \nonumber \\
&+& \frac{1}{2} \left[ \delta \Pi_I^2
+ \frac{\delta \chi_R^2}{A_0 \psi_0^4} + Q(r) \delta \varphi_R^2 \right] \; .
\end{eqnarray}
Notice now that, for any perturbation that falls into the star from
outside, the energy density, and hence the total mass, will
necessarily increase, as the linear terms in the expression above
all vanish and we will be left with a positive definite contribution from
the quadratic terms.

However, for internal perturbations we can again use the condition
$\delta \Pi_I = - (\omega /\alpha_0) \: \delta \varphi_R$ introduced above
to find:
\begin{eqnarray}
\rho_E &=& (\rho_E)_0 + \varphi_0 \left( Q(r) - \frac{\omega^2}{\alpha_0^2} \right) \delta \varphi_R
+ \frac{\varphi_0' \delta \chi_R}{A_0 \psi_0^4} \nonumber \\
&+& \frac{1}{2} \left[ \left( Q(r) + \frac{\omega^2}{\alpha_0^2} \right) \delta \varphi_R^2
+ \frac{\delta \chi_R^2}{A_0 \psi_0^4} \right] \; ,
\end{eqnarray}
where we also already substituted $(\Pi_I)_0 = \omega \varphi_0 /
\alpha_0$.  For small perturbations the linear contribution dominates,
and it does not have a definite sign, so the total mass of the
spacetime can increase or decrease.

In summary, we will consider three different types of perturbations
for the simulations presented below, all of which will be such that
$\delta \Pi_R = \delta \varphi_I = \delta \chi_I = 0$ (so that the
initial momentum density vanishes).

\begin{itemize}

\item{TYPE I:} An {\em internal} perturbation such that \mbox{$\delta
  \varphi_R \neq 0$} and $\delta \Pi_I = 0$.  This perturbation
  changes the boson density.

\item{TYPE II:} An {\em internal} perturbation such that \mbox{$\delta
  \Pi_I = - (\omega /\alpha_0) \: \delta \varphi_R$}. This perturbation
  preserves the boson density to linear order in small quantities, and
  can either increase or decrease the total mass of the star.
  Interestingly, in practice we have found that these type of
  perturbations also seem to have a very small effect on the value of
  the total mass.

\item{TYPE III:} An {\em external} perturbation (scalar field falling
  into the star from outside) with \mbox{$\delta \Pi_I = \pm (\omega /
    \alpha_0) \: \delta \varphi_R$}, which again preserves the boson
  density to linear order in small quantities but always increases the
  mass.

\end{itemize}

\noindent Notice that in all three cases we have $\delta \varphi_R$ as a
free parameter.

\vspace{5mm}

Finally, in order to find the perturbed initial data we choose values
of $\ell$ and $\varphi_0$, and solve for the unperturbed configuration
first. Having found the functions $\varphi(r)$, $A(r)$ and $\alpha(r)$
and the frequency $\omega$ for the unperturbed case, we add small
perturbations to $\varphi(r)$ and $\Pi_I(r)$ corresponding to one of
the three types described above (remember that for the unperturbed
case we have $(\Pi_I)_0 = \omega \varphi_0 / \alpha_0$), and solve
again the Hamiltonian constraint to find the modified value of $A(r)$.
We also solve again the polar slicing condition for a new lapse
$\alpha(r)$, which will differ slightly from its unperturbed value
$\alpha_0(r)$. This guarantees that the perturbed configuration will
still be such that $\partial_t K_{\theta \theta}=0$ initially (but
this will not remain so at later times, as perturbed configurations
are no longer stationary).


\section{Analysis tools and numerical code}
\label{sec:numerical}


\subsection{Gauge choice}
\label{sec:gauge}

For our simulations we choose for simplicity a vanishing shift, and
for the lapse function we choose the standard ``1+log'' slicing
condition, which has the form~\cite{Alcubierre08a}:
\begin{equation}
\partial_t \alpha = - 2 \alpha \:  K \; ,
\end{equation}
where $\alpha$ is the lapse function and $K=K^m_m$ the trace of the
extrinsic curvature.  This condition is very robust in practice and
allows for long-lived and stable evolutions.

Notice that the initial data, both in the perturbed and unperturbed
case, is such that $K(t=0)=0$ (in fact the whole extrinsic curvature
vanishes).  In the unperturbed case the 1+log slicing condition
should guarantee that the lapse remains static up to numerical
truncation error.  For the perturbed cases, however, we expect $K$ to
evolve away from 0 from the beginning, resulting also in a dynamical
lapse.


\subsection{Total mass, boson number and binding energy}
\label{sec:masscharge}

As already mentioned, the Eulerian observers measure an energy and
boson density given by equations~\eqref{eq:Edensity}
and~\eqref{eq:Bdensity} above. These quantities can be used to define
a total mass and conserved boson (particle) number.

For the total mass we go back to the Hamiltonian constraint, which in
general has the form:
\begin{equation}
R + K^2 - K_{ij} K^{ij} = 16 \pi \rho_E \; ,
\end{equation}
with $R$ the three-dimensional Ricci scalar.  Now, in spherical
symmetry, and using the areal radius $r_a^2$, the spatial metric can be
written as:
\begin{equation}
dl^2 = \frac{dr_a^2}{1 - 2 m(r_a)/r_a} + r_a^2 d \Omega^2 \; ,
\label{eq:metricareal}
\end{equation}
with $m(r_a)$ the so-called ``Misner-Sharp mass function''~\cite{cMdS64}. In these
coordinates the Ricci scalar becomes:
\begin{equation}
R = \left( \frac{4}{r_a^2} \right) \frac{dm}{d r_a} \; ,
\end{equation}
so the Hamiltonian constraint implies:
\begin{equation}
\frac{dm}{d r_a} = r_a^2 \left[ 4 \pi \rho_E + \frac{1}{4} \left( K_{ij} K^{ij} - K^2 \right) \right] \: .
\end{equation}
The mass function can then be integrated to define a total mass $M$ as:
\begin{equation}
M := \int_0^\infty \left[ 4 \pi \rho_E + \frac{1}{4} \left( K_{ij} K^{ij} - K^2 \right) \right] r_a^2 dr_a \; ,
\end{equation}
where $K_{ij} K^{ij} = (K^r_r)^2 + 2 (K^\theta_\theta)^2$.  Notice
that if $K_{ij}=0$ the above expression is essentially identical to
the Newtonian definition of mass (but we need to be in the areal
gauge). Now, if the sources have compact support (or decay
exponentially), the spacetime will reduce to Schwarzschild far away,
and $M$ will correspond to the total
ADM mass of the system.

On the other hand, the areal radius is given in terms of our
coordinate radius $r$ as $r_a = r \psi^2 B^{1/2}$
(confront~\eqref{eq:metricareal} with~\eqref{eq:metric}), which
implies:
\begin{equation}
r_a^2 dr_a = r^2 \psi^6 B^{3/2} \left[ 1 + r \left( \frac{\partial_r B}{2B}
+ 2 \frac{\partial_r \psi}{\psi} \right) \right] dr \; .
\end{equation}
The final expression for the total mass is then:
\begin{eqnarray}
M &:=& \int_0^\infty r^2 \psi^6 B^{3/2} \left[ 4 \pi \rho_E
+ \frac{1}{4} \left( K_{ij} K^{ij} - K^2 \right) \right] \nonumber \\
&& \times  \left[ 1 + r \left( \frac{\partial_r B}{2B}
+ 2 \frac{\partial_r \psi}{\psi} \right) \right] dr  \; ,
\label{eq:totalmass}
\end{eqnarray}
This expression is valid for any spherically symmetric metric
parametrized as in Eq.~\eqref{eq:metric}.

Let us turn now to the total boson number. For a complex scalar field
it is well known that there exists a conserved current particle
$J^\mu$ such that $\nabla_\mu J^\mu=0$ (see Eq.~\eqref{eq:Jcurrent}
above). This immediately implies that the integral of the boson
density $\rho_B = - n^\mu J_\mu$ is a conserved quantity, which we
refer to as the ``total boson number'' $N_B$:
\begin{eqnarray}
N_B &:=& \int \rho_B \gamma^{1/2} dr d\theta d\varphi \nonumber \\
 &=& 4 \pi \int_0^\infty \rho_B \left( A^{1/2} B
\psi^6 \right) r^2 dr \; ,
\label{eq:bosonnumber}
\end{eqnarray}
with $\gamma=AB^2 \psi^{12}$ the determinant of the spatial metric.
Notice that if the boson particles associated with the complex scalar
fields had an electric charge $q$, the total charge would simply be
$Q=q N_B$.

\vspace{5mm}

One last concept that needs to be introduced is that of ``binding
energy''.  The binding energy $U$ is a measure of the difference
between the total mass-energy of the system, given by the ADM mass
$M$, and the rest mass of the bosons, which can be simply defined as
$\mu N_B$, with $\mu$ the mass of the scalar field:
\begin{equation}
U := M - \mu N_B \; .
\label{eq:binding}
\end{equation}
If the binding energy is negative, we should have a bound gravitational
system, while if it is positive the system is unbound.


\subsection{Apparent horizons and horizon mass}
\label{sec:horizon}

As we will see below, when we perturb boson stars in the unstable
branch they can collapse to form a black hole.  We identify the
presence of such a black hole by looking for the appearance of an
apparent horizon, that is the outermost closed two-surface where the
expansion of outgoing null geodesics vanishes.  In the case of
spherical symmetry this is rather straightforward, and reduces to the
following condition~\cite{Alcubierre08a}:
\begin{equation}
\frac{1}{\psi^2 \sqrt{A}} \left( \frac{2}{r} + \frac{\partial_r B}{B}
+ 4 \: \frac{\partial \psi}{\psi} \right) - 2 K^\theta_\theta = 0 \; .
\end{equation}
Notice that the above equation should not be understood as a
differential equation, but rather as an algebraic condition that, when
satisfied for some value of $r$, indicates the presence of an apparent
horizon at that location. If the condition is satisfied at more than
one place, the apparent horizon will correspond to the outermost
location.

Once we have located an apparent horizon at some coordinate radius
$r=r_H$ we can calculate its area as \mbox{$A_{\rm H} = 4 \pi r_a^2 = 4
  \pi r_H^2 \psi_H^4 B_H$}, with $r_a$ the areal radius as before, and
from there obtain the so-called ``horizon mass'' as follows:
\begin{equation}
M_{\rm H}= \sqrt{ \frac{A_{\rm H}}{16\pi} } = \frac{r_H \psi_H^2 B_H^{1/2}}{2}  \; .
\label{eq:BHmass}
\end{equation}
This horizon mass should always be smaller than, or equal to, the
total ADM mass $M$ of the spacetime.


\subsection{Numerical code}
\label{sec:numerics}

Our simulations are carried out with the {\em OllinSphere} code, a
generic numerical relativity finite-difference code for spherical
symmetry.  The initial data is obtained using a shooting method with
fourth order Runge-Kutta on a regular grid. Our grid staggers the
origin to avoid having divisions by zero for terms of type $1/r$.

For the evolution we use a BSSN formulation adapted to spherical
symmetry~\cite{Alcubierre:2010is}. The code uses a method of lines
with fourth order spatial differences, and a fourth order Runge-Kutta
time integrator.  This code has been previously tested with real
scalar fields, and has been used in the context of scalar-tensor
theories of gravity with minimal
modifications~\cite{Alcubierre:2010ea,Ruiz:2012jt}. The exterior
boundary conditions are of a constraint-preserving type, following the
method described in~\cite{Alcubierre:2014joa}.


\section{Numerical simulations}
\label{sec:results}


\subsection{General considerations}
\label{sec:num_considerations}

We have performed a series of dynamical simulations for $\ell$-boson
stars, for different value of $\ell$ in the range $\ell=0,1,2,3,4$.
In each case we have performed simulations of both the unperturbed
solutions, and different perturbations of the three types discussed
above in Section~\ref{sec:perturbations}.  In all cases considered
here we have chosen for simplicity the boson mass equal to unity,
$\mu=1$.

Before going into our results, there are several properties of the
$\ell$-boson stars that need to be discussed.  As mentioned above, for
a fixed value of $\ell$ the scalar field close to the origin behaves
as $\varphi(r) \sim \varphi_0 r^\ell$, and for each value of
$\varphi_0$ one needs to solve an eigenvalue problem to find the
oscillation frequency $\omega$. Parametrizing the solutions for each
$\ell$ with $\varphi_0$, one finds that as we increase $\varphi_0$ the
ADM mass $M$ of the configurations first increases and reaches a
maximum, after which it decreases again. These results where already
presented in~\cite{Alcubierre:2018ahf}.

In that reference, however, we did not compute the total boson number
$N_B$ and binding energy $U$ for each solution.  Doing that we find
that the boson number increases with $\varphi_0$ until it reaches a
maximum at the same point as the total mass $M$, and then also
decreases. The binding energy $U$, on the other hand, starts negative
and decreases, until it reaches a minimum just as the mass and boson
number reach a maximum. It then starts to increase and at some point
becomes positive, corresponding to solutions that are no longer
gravitationally bound.

\begin{figure}
\includegraphics[width=0.49\textwidth]{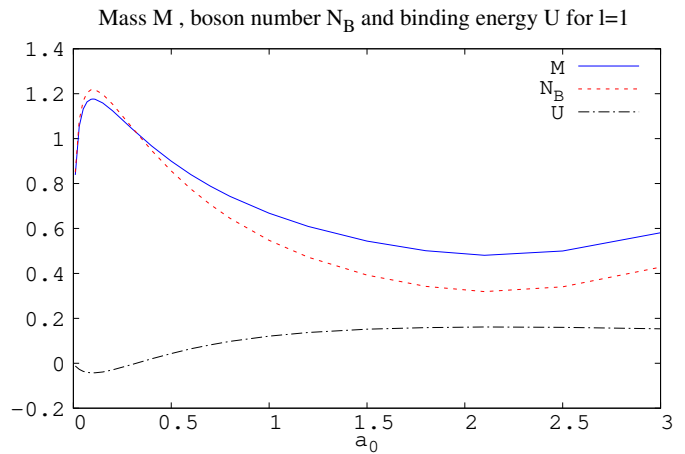}  
\caption{Total ADM mass $M$, total boson number $N_B$, and binding
  energy $U$ for the case for $\ell=1$ and $\mu=1$. The configurations
  here are parametrized with $a_0$, which is given in terms of
  $\varphi_0$ as \mbox{$a_0 = [4 \pi (2 \ell+1)]^{1/2} \varphi_0$}.}
\label{fig:l=1-static}
\end{figure}

Figure~\ref{fig:l=1-static} shows a plot of the total ADM mass $M$,
total boson number $N_B$, and binding energy $U$ for the case with
$\ell=1$.  The configurations here are parametrized with $a_0$, which
is given in terms of $\varphi_0$ as \mbox{$a_0 = [4 \pi (2
    \ell+1)]^{1/2} \varphi_0$} (this is in order to be consistent with
the normalization used in~\cite{Alcubierre:2018ahf}). Notice how there
is a region where $M<N_B$ corresponding to bound configurations, and a
region with $M>N_B$ corresponding to unbound states.  Similar plots
can be found for other values of $\ell$.

For each value of $\ell$, there are then three regions of interest in
the parameter space of solutions: stable bound configurations with
$M<M_{\rm max}$ and $M<N_B$, unstable bound configurations with
$M>M_{\rm max}$ and $M<N_B$, and finally unstable unbound
configurations with $M>M_{\rm max}$ and $M>N_B$.


\subsection{Summary of results}
\label{sec:num_summary}

As mentioned above, for each value of $\ell$ there are three regions
of interest in parameter space: stable configurations, unstable bound
configurations, and unstable unbound configurations. Let us denote by
$\varphi_0^M$ the value of the parameter $\varphi_0$ for which we
obtain the maximum ADM mass, and by $\varphi_0^U$ the value for which
the binding energy is zero.  We find that in general $\varphi_0^M <
\varphi_0^U$.

For all values of $\ell$ we have studied, the results of our
simulations can be summarized as follows:

\begin{itemize}

\item The region $0 < \varphi_0 < \varphi_0^M$ corresponds to bound
  stable configurations. For all types of (small) perturbations
  studied, these configurations oscillate around the stationary
  solution.  The oscillations are extremely long-lived, but they seem
  to slowly settle down to a stationary solution that lies close to the
  original one.

\item The region $\varphi_0^M<\varphi_0 < \varphi_0^U$ corresponds to
  unstable but bound configurations that, depending on the specific
  type of perturbation, can either collapse to form a black hole or
  ``migrate'' to the stable branch. This migration to the stable
  branch is achieved by ejecting excess scalar field to
  infinity. Again, these migrating solutions in fact oscillate for
  extremely long times and seem to very slowly settle down to a
  stationary solution.

\item The region $\varphi_0>\varphi_0^U$ corresponds to unstable and
  unbound solutions that, depending on the specific type of
  perturbation, can either collapse to a black hole or dissipate
  (``explode'') to infinity.  Dissipating solutions may oscillate a
  few times before they dissipate completely.

\end{itemize}

For standard boson stars with $\ell=0$, the difference in behaviour
between bound and unbound unstable configurations that do not collapse
to form a black hole, that is configurations that either migrate to
the stable branch or dissipate to infinity, has already been
observed~\cite{Balakrishna:1997ej,Guzman04,Guzman09}. Interestingly, in one of the
original papers on perturbed $\ell=0$ boson stars by Seidel and
Suen~\cite{Seidel90}, the authors mention that they observe no
difference between unstable configurations with negative or positive
binding energy.  This could be related to the specific types of
perturbations they studied.

Tables~\ref{table:l=0} to~\ref{table:l=4} present results from a
battery of simulations we have performed for values of $\ell$ in the
range $\ell=0,1,2,3,4$.  In each case, we have considered all three
types of perturbations described in Section~\ref{sec:perturbations}
above. We have added also perturbations of ``type 0'', which in fact
correspond to evolutions of the unperturbed initial data.  Notice that
these ``unperturbed'' evolutions are in fact slightly perturbed by
numerical truncation error.

\begin{figure}

\begin{overpic}[width=0.49\textwidth]{stability.eps}  

\put(89,64.5){\includegraphics[scale=0.008]{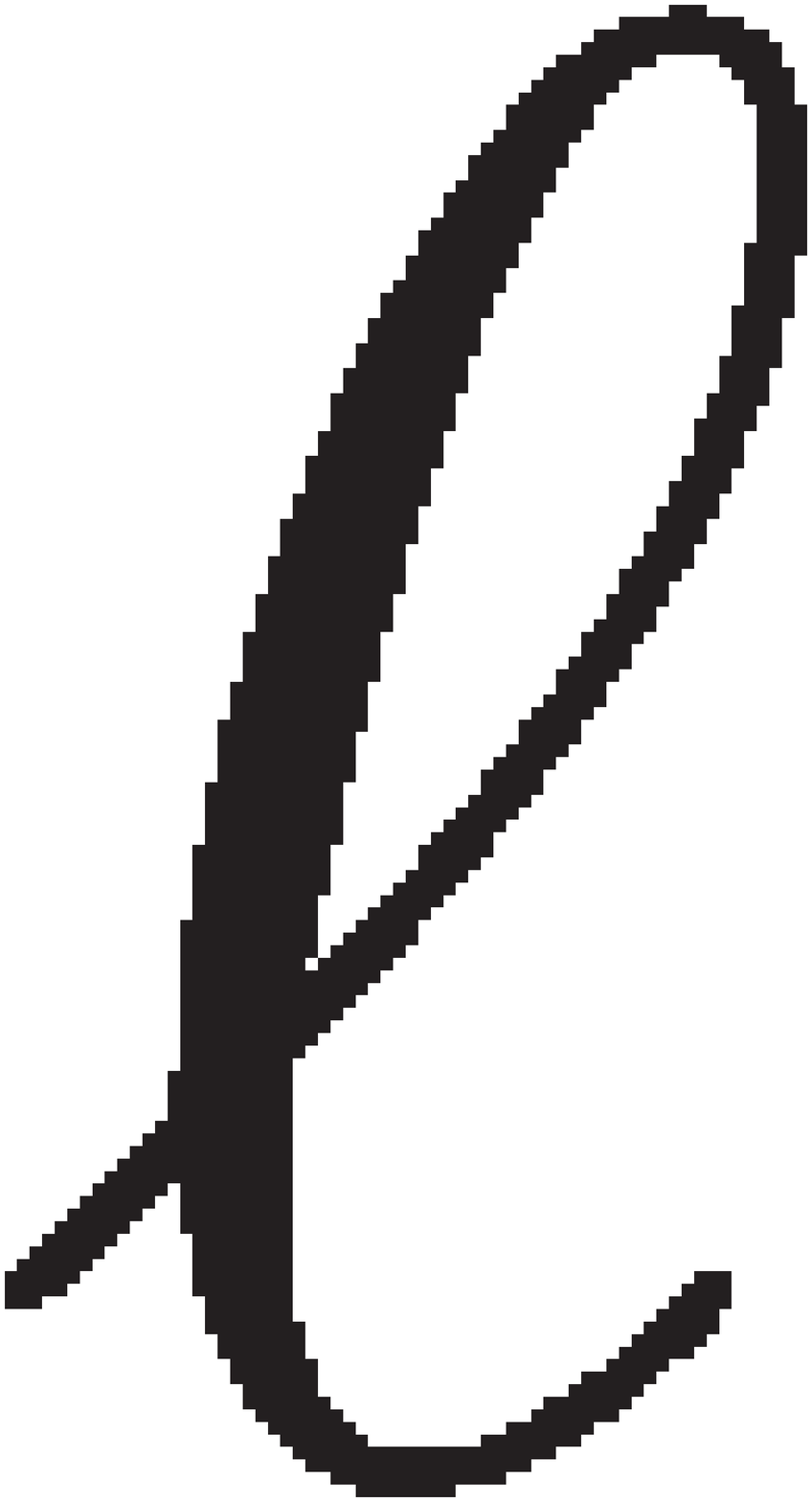}}
\put(89,61.3){\includegraphics[scale=0.008]{ell.eps}}
\put(89,58.1){\includegraphics[scale=0.008]{ell.eps}}
\put(89,54.9){\includegraphics[scale=0.008]{ell.eps}}
\put(89,51.7){\includegraphics[scale=0.008]{ell.eps}}

\end{overpic}
\caption{The three regions of stability for $\ell$-boson stars. See
  the text for a detailed explanation.}
\label{fig:summary}
\end{figure}

Figure~\ref{fig:summary} shows the three regions of stability for
\mbox{$\ell=0,1,2,3,4$}. The figure shows a plot of the mass of the
configuration $M$ as a function of the oscillation frequency
$\omega$. Configurations to the right of the maximum mass line (which
coincides with the minimum binding energy) correspond to bound stable
configurations. The diamonds indicate those specific stable
configurations that where evolved.  The central region corresponds to
unstable but bound configurations.  Squares represent those specific
configurations that were evolved in this region, and either collapse
to a black hole or migrate to a stable configuration.  Finally, all
those configurations to the left of the $U=0$ line (zero binding
energy) correspond to unstable and unbound solutions. The triangles
correspond to those specific configurations that we evolved, and either
collapse to a black hole or disperse to infinity.

\vspace{3mm}

In all our simulations, we used a small Gaussian perturbation to the
scalar field $\delta \varphi_R$:
\begin{equation}
\delta \varphi_R(r) = \epsilon \; \exp \left[ -(r-r_0)/\sigma^2 \right] \; ,
\end{equation}
with $\epsilon$ the amplitude of the perturbation and $\sigma$ its
width.  When the perturbations are internal to the star (types I and
II), we choose $r_0$ to coincide roughly with the place where the
scalar field $\varphi(r)$ has a maximum (notice that for
\mbox{$\ell>0$} this maximum is not at the origin).  The amplitude of
the perturbation is rescaled with this maximum, and for simplicity we
always take the width of the Gaussian to be equal to unity,
$\sigma=1$. For the perturbation of the imaginary part of the time
derivative of the scalar field, $\Pi_I(r)$, we take
\begin{equation}
\delta \Pi_I(r) = s \: (\omega / \alpha) \: \delta \varphi_R(r) ,
\end{equation}
with $s=0$ for perturbations of type I, $s = -1$ for
type II, and \mbox{$s = \pm 1$} for type III.

In all the simulations described here we have taken a grid spacing of
$\Delta r=0.02$ and a total of $2500$ grid points, so the outer
boundary is located at $r=49.99$ (remember that we stagger the
origin). For the time stepping we take $\Delta t=0.01$, and we evolve
for $50,000$ time steps, corresponding to a final time $t=500$.  We
have in fact performed simulations with different grid spacings to
verify fourth order convergence, and also much longer simulations in
some special cases to study the late time behaviour of solutions that
migrate to the stable branch or explode to infinity (see
Section~\ref{sec:num_examples} below). The main effect of using a
higher resolution is that those perturbations of type 0 that collapse
to a black hole do so at later times for higher resolution runs.  This
is to be expected since in that case the perturbation is only through
numerical truncation error which is smaller for higher resolution.

From the tables one can see some interesting facts. First, for all
types of (small) perturbations with \mbox{$0 < \varphi_0 <
  \varphi_0^M$}, and all values of $\ell$, the configurations are
stable as expected.  In the region $\varphi_0^M<\varphi_0 <
\varphi_0^U$, the configurations are unstable and either collapse to a
black hole or migrate to the stable branch.  But collapse to a black
hole is far more common, and we find that only type I perturbations
with $\epsilon<0$, or type II perturbations with $\epsilon>0$ can
migrate to the stable branch. Moreover, for type II perturbations with
\mbox{$\epsilon>0$}, migration to the stable branch only happens for
very small values of $\epsilon$, and increasing slightly the
perturbation amplitude again results in collapse to a black hole. The
transition between migration and collapse for these type of
perturbations seems to be related not so much with the sign of the
binding energy $U$, which in these region is always negative, but
rather with the value of $dU/d\epsilon$ (that is, if $U$ is decreasing
or increasing with $\epsilon$), but this still needs more
studying. Finally, in the region $\varphi_0 > \varphi_0^U$ the
configurations are also unstable and either collapse to a black hole
of explode to infinity.  Again, collapse is far more common and only
type I perturbations with $\epsilon<0$, or type II perturbations with
$\epsilon>0$ (and very small) explode to infinity.

Interestingly, for type 0 perturbations in the unstable branch
$\varphi_0 > \varphi_0^M$, we always find collapse to a black hole
except for one particular case with $\ell=3$ for which the
configuration migrates to the stable branch.  Of course, these
perturbations are only through numerical truncation error which we can
not control.

\begin{table*}
\begin{tabular}{|c|c|c|c|c|c|c|c|c|c|c|}
\hline
$\ell$ & $a_0$ & $\omega$& Perturbation & $M$ & $N_B$ & $U$ & 
$\epsilon/ \varphi^{\rm max}_R $ & $s$ & $r_0$ & End result \\
\hline \hline
0 & 0.2 & 0.88401 & Type 0   & 0.6209 & 0.6391 & $-0.0182$ & --       & --   &  --  & stable \\
0 & 0.2 & 0.88401 & Type I   & 0.6211 & 0.6394 & $-0.0183$ & $+0.005$ & 0    &  0.0 & stable \\
0 & 0.2 & 0.88401 & Type I   & 0.6207 & 0.6389 & $-0.0182$ & $-0.005$ & 0    &  0.0 & stable \\
0 & 0.2 & 0.88401 & Type II  & 0.6209 & 0.6391 & $-0.0182$ & $+0.005$ & $-1$ &  0.0 & stable \\
0 & 0.2 & 0.88401 & Type II  & 0.6209 & 0.6391 & $-0.0182$ & $-0.005$ & $-1$ &  0.0 & stable \\
0 & 0.2 & 0.88401 & Type III & 0.6238 & 0.6412 & $-0.0174$ & $+0.01$  & $+1$ & 20.0 & stable \\
0 & 0.2 & 0.88401 & Type III & 0.6237 & 0.6372 & $-0.0135$ & $+0.01$  & $-1$ & 20.0 & stable \\
0 & 0.4 & 0.80866 & Type 0   & 0.6088 & 0.6235 & $-0.0147$ & --       & --   &  --  & black hole \\
0 & 0.4 & 0.80866 & Type I   & 0.6096 & 0.6246 & $-0.0150$ & $+0.005$ & 0    &  0.0 & black hole \\
0 & 0.4 & 0.80866 & Type I   & 0.6079 & 0.6225 & $-0.0146$ & $-0.005$ & 0    &  0.0 & migration to stable branch \\
0 & 0.4 & 0.80866 & Type II  & 0.6087 & 0.6235 & $-0.0148$ & $+0.005$ & $-1$ &  0.0 & migration to stable branch \\
0 & 0.4 & 0.80866 & Type II  & 0.6088 & 0.6236 & $-0.0148$ & $-0.005$ & $-1$ &  0.0 & black hole \\
0 & 0.4 & 0.80866 & Type III & 0.6193 & 0.6305 & $-0.0112$ & $+0.01$  & $+1$ & 20.0 & black hole \\
0 & 0.4 & 0.80866 & Type III & 0.6193 & 0.6166 & $+0.0027$ & $+0.01$  & $-1$ & 20.0 & black hole \\
0 & 0.6 & 0.77134 & Type 0   & 0.5248 & 0.5167 & $+0.0081$ & --       & --   &  --  & black hole \\
0 & 0.6 & 0.77134 & Type I   & 0.5266 & 0.5190 & $+0.0075$ & $+0.005$ & 0    &  0.0 & black hole \\
0 & 0.6 & 0.77134 & Type I   & 0.5230 & 0.5144 & $+0.0086$ & $-0.005$ & 0    &  0.0 & explosion to infinity \\
0 & 0.6 & 0.77134 & Type II  & 0.5246 & 0.5165 & $+0.0081$ & $+0.005$ & $-1$ &  0.0 & explosion to infinity \\
0 & 0.6 & 0.77134 & Type II  & 0.5250 & 0.5169 & $+0.0081$ & $-0.005$ & $-1$ &  0.0 & black hole \\
0 & 0.6 & 0.77134 & Type III & 0.5481 & 0.5314 & $+0.0167$ & $+0.01$  & $+1$ & 20.0 & black hole \\
0 & 0.6 & 0.77134 & Type III & 0.5480 & 0.5020 & $+0.0460$ & $+0.01$  & $-1$ & 20.0 & black hole \\
\hline
\end{tabular}
\caption{Results of simulations for $\ell=0$. Notice that in this case
  the maximum mass is $M_{\rm max} \simeq 0.633$.}
\label{table:l=0}
\end{table*}

\begin{table*}
\begin{tabular}{|c|c|c|c|c|c|c|c|c|c|c|}
\hline
$\ell$ & $a_0$ & $\omega$& Perturbation & $M$ & $N_B$ & $U$ & 
$\epsilon/ \varphi^{\rm max}_R $ & $s$ & $r_0$ & End result \\
\hline \hline
1 & 0.05 & 0.88253 & Type 0   & 1.1316 & 1.1674 & $-0.0358$ & --       & --   &  --  & stable \\
1 & 0.05 & 0.88253 & Type I   & 1.1344 & 1.1705 & $-0.0361$ & $+0.01$  & 0    &  4.5 & stable \\
1 & 0.05 & 0.88253 & Type I   & 1.1288 & 1.1642 & $-0.0354$ & $-0.01$  & 0    &  4.5 & stable \\
1 & 0.05 & 0.88253 & Type II  & 1.1316 & 1.1673 & $-0.0357$ & $+0.01$  & $-1$ &  4.5 & stable \\
1 & 0.05 & 0.88253 & Type II  & 1.1316 & 1.1673 & $-0.0357$ & $-0.01$  & $-1$ &  4.5 & stable \\
1 & 0.05 & 0.88253 & Type III & 1.1324 & 1.1681 & $-0.0357$ & $+0.01$  & $+1$ & 20.0 & stable \\
1 & 0.05 & 0.88253 & Type III & 1.1320 & 1.1670 & $-0.0350$ & $+0.01$  & $-1$ & 20.0 & stable \\
1 & 0.2  & 0.78330 & Type 0   & 1.1241 & 1.1536 & $-0.0295$ & --       & --   &  --  & black hole \\
1 & 0.2  & 0.78330 & Type I   & 1.1297 & 1.1608 & $-0.0311$ & $+0.01$  & 0    &  2.5 & black hole \\
1 & 0.2  & 0.78330 & Type I   & 1.1185 & 1.1465 & $-0.0280$ & $-0.01$  & 0    &  2.5 & migration to stable branch \\
1 & 0.2  & 0.78330 & Type II  & 1.1240 & 1.1536 & $-0.0296$ & $+0.005$ & $-1$ &  2.5 & migration to stable branch \\
1 & 0.2  & 0.78330 & Type II  & 1.1241 & 1.1535 & $-0.0294$ & $+0.01$  & $-1$ &  2.5 & black hole \\
1 & 0.2  & 0.78330 & Type II  & 1.1242 & 1.1537 & $-0.0295$ & $-0.01$  & $-1$ &  2.5 & black hole \\
1 & 0.2  & 0.78330 & Type III & 1.1261 & 1.1550 & $-0.0289$ & $+0.01$  & $+1$ & 20.0 & black hole \\
1 & 0.2  & 0.78330 & Type III & 1.1261 & 1.1522 & $-0.0261$ & $+0.01$  & $-1$ & 20.0 & black hole \\
1 & 0.4  & 0.74471 & Type 0   & 0.9674 & 0.9476 & $+0.0198$ & --       & --   &  --  & black hole \\
1 & 0.4  & 0.74471 & Type I   & 0.9743 & 0.9568 & $+0.0175$ & $+0.01$  & 0    &  1.7 & black hole \\
1 & 0.4  & 0.74471 & Type I   & 0.9606 & 0.9385 & $+0.0221$ & $-0.01$  & 0    &  1.7 & explosion to infinity \\
1 & 0.4  & 0.74471 & Type II  & 0.9673 & 0.9473 & $+0.0200$ & $+0.01$  & $-1$ &  1.7 & explosion to infinity \\
1 & 0.4  & 0.74471 & Type II  & 0.9677 & 0.9478 & $+0.0199$ & $-0.01$  & $-1$ &  1.7 & black hole \\
1 & 0.4  & 0.74471 & Type III & 0.9714 & 0.9502 & $+0.0212$ & $+0.01$  & $+1$ & 20.0 & black hole \\
1 & 0.4  & 0.74471 & Type III & 0.9714 & 0.9450 & $+0.0264$ & $+0.01$  & $-1$ & 20.0 & black hole \\
\hline
\end{tabular}
\caption{Results of simulations for $\ell=1$. Notice that in this case
  the maximum mass is $M_{\rm max} \simeq 1.176$.}
\label{table:l=1}
\end{table*}

\begin{table*}
\begin{tabular}{|c|c|c|c|c|c|c|c|c|c|c|}
\hline
$\ell$ & $a_0$ & $\omega$& Perturbation & $M$ & $N_B$ & $U$ & 
$\epsilon/ \varphi^{\rm max}_R $ & $s$ & $r_0$ & End result \\
\hline \hline
2 & 0.005 & 0.88354 & Type 0   & 1.6268 & 1.6793 & $-0.0525$ & --       & --   &  --  & stable \\
2 & 0.005 & 0.88354 & Type I   & 1.6307 & 1.6837 & $-0.0530$ & $+0.01$  & 0    &  8.0 & stable \\
2 & 0.005 & 0.88354 & Type I   & 1.6229 & 1.6749 & $-0.0520$ & $-0.01$  & 0    &  8.0 & stable \\
2 & 0.005 & 0.88354 & (A) Type II  & 1.6268 & 1.6792 & $-0.0524$ & $+0.01$  & $-1$ &  8.0 & stable \\
2 & 0.005 & 0.88354 & Type II  & 1.6268 & 1.6793 & $-0.0525$ & $-0.01$  & $-1$ &  8.0 & stable \\
2 & 0.005 & 0.88354 & Type III & 1.6273 & 1.6797 & $-0.0524$ & $+0.01$  & $+1$ & 30.0 & stable \\
2 & 0.005 & 0.88354 & Type III & 1.6273 & 1.6789 & $-0.0516$ & $+0.01$  & $-1$ & 30.0 & stable \\
2 & 0.05  & 0.76114 & Type 0   & 1.6035 & 1.6388 & $-0.0353$ & --       & --   &  --  & black hole \\
2 & 0.05  & 0.76114 & Type I   & 1.6121 & 1.6502 & $-0.0381$ & $+0.01$  & 0    &  4.0 & black hole  \\
2 & 0.05  & 0.76114 & (B) Type I   & 1.5949 & 1.6276 & $-0.0327$ & $-0.01$  & 0    &  4.0 & migration to stable branch \\
2 & 0.05  & 0.76114 & Type II  & 1.6035 & 1.6388 & $-0.0353$ & $+0.005$ & $-1$ &  4.0 & migration to stable branch \\
2 & 0.05  & 0.76114 & Type II  & 1.6035 & 1.6387 & $-0.0352$ & $+0.01$  & $-1$ &  4.0 & black hole \\
2 & 0.05  & 0.76114 & Type II  & 1.6036 & 1.6389 & $-0.0353$ & $-0.01$  & $-1$ &  4.0 & black hole \\
2 & 0.05  & 0.76114 & Type III & 1.6062 & 1.6407 & $-0.0345$ & $+0.01$  & $+1$ & 30.0 & black hole \\
2 & 0.05  & 0.76114 & Type III & 1.6062 & 1.6370 & $-0.0308$ & $+0.01$  & $-1$ & 30.0 & black hole \\
2 & 0.1   & 0.73427 & Type 0   & 1.4424 & 1.4231 & $+0.0193$ & --       & --   &  --  & black hole \\
2 & 0.1   & 0.73427 & Type I   & 1.4521 & 1.4363 & $+0.0158$ & $+0.01$  & 0    &  3.0 & black hole \\
2 & 0.1   & 0.73427 & Type I   & 1.4328 & 1.4100 & $+0.0228$ & $-0.01$  & 0    &  3.0 & explosion to infinity \\
2 & 0.1   & 0.73427 & (C) Type II  & 1.4424 & 1.4230 & $+0.0194$ & $+0.005$ & $-1$ &  3.0 & explosion to infinity \\
2 & 0.1   & 0.73427 & Type II  & 1.4424 & 1.4229 & $+0.0195$ & $+0.01$  & $-1$ &  3.0 & black hole \\
2 & 0.1   & 0.73427 & Type II  & 1.4426 & 1.4232 & $+0.0194$ & $-0.01$  & $-1$ &  3.0 & black hole \\
2 & 0.1   & 0.73427 & (D) Type III & 1.4466 & 1.4258 & $+0.0208$ & $+0.01$  & $+1$ & 30.0 & black hole \\
2 & 0.1   & 0.73427 & Type III & 1.4466 & 1.4203 & $+0.0263$ & $+0.01$  & $-1$ & 30.0 & black hole \\
\hline
\end{tabular}
\caption{Results of simulations for $\ell=2$. Notice that in this case
  the maximum mass is $M_{\rm max} \simeq 1.714$.}
\label{table:l=2}
\end{table*}

\begin{table*}
\begin{tabular}{|c|c|c|c|c|c|c|c|c|c|c|}
\hline
$\ell$ & $a_0$ & $\omega$& Perturbation & $M$ & $N_B$ & $U$ & 
$\epsilon/ \varphi^{\rm max}_R $ & $s$ & $r_0$ & End result \\
\hline \hline
3 & 0.0005 & 0.86561 & Type 0   & 2.1782 & 2.2560 & $-0.0778$ & --       & --   &  --  & stable \\
3 & 0.0005 & 0.86561 & Type I   & 2.1835 & 2.2621 & $-0.0786$ & $+0.01$  & 0    & 10.0 & stable \\
3 & 0.0005 & 0.86561 & Type I   & 2.1729 & 2.2500 & $-0.0771$ & $-0.01$  & 0    & 10.0 & stable \\
3 & 0.0005 & 0.86561 & Type II  & 2.1782 & 2.2560 & $-0.0778$ & $+0.01$  & $-1$ & 10.0 & stable \\
3 & 0.0005 & 0.86561 & Type II  & 2.1782 & 2.2560 & $-0.0778$ & $-0.01$  & $-1$ & 10.0 & stable \\
3 & 0.0005 & 0.86561 & Type III & 2.1787 & 2.2564 & $-0.0777$ & $+0.01$  & $+1$ & 30.0 & stable \\
3 & 0.0005 & 0.86561 & Type III & 2.1786 & 2.2558 & $-0.0772$ & $+0.01$  & $-1$ & 30.0 & stable \\
3 & 0.005  & 0.76579 & Type 0   & 2.1519 & 2.2159 & $-0.0640$ & --       & --   &  --  & migration to stable branch \\
3 & 0.005  & 0.76579 & Type I   & 2.1621 & 2.2291 & $-0.0670$ & $+0.01$  & 0    &  6.0 & black hole \\
3 & 0.005  & 0.76579 & Type I   & 2.1419 & 2.2028 & $-0.0609$ & $-0.01$  & 0    &  6.0 & migration to stable branch \\
3 & 0.005  & 0.76579 & Type II  & 2.1519 & 2.2159 & $-0.0640$ & $+0.002$ & $-1$ &  6.0 & migration to stable branch \\
3 & 0.005  & 0.76579 & Type II  & 2.1520 & 2.2158 & $-0.0638$ & $+0.01$  & $-1$ &  6.0 & black hole \\
3 & 0.005  & 0.76579 & Type II  & 2.1521 & 2.2159 & $-0.0638$ & $-0.01$  & $-1$ &  6.0 & black hole \\
3 & 0.005  & 0.76579 & Type III & 2.1534 & 2.2170 & $-0.0636$ & $+0.01$  & $+1$ & 30.0 & black hole \\
3 & 0.005  & 0.76579 & Type III & 2.1534 & 2.2149 & $-0.0615$ & $+0.01$  & $-1$ & 30.0 & black hole \\
3 & 0.02   & 0.72405 & Type 0   & 1.8432 & 1.7997 & $+0.0435$ & --       & --   &  --  & black hole \\
3 & 0.02   & 0.72405 & Type I   & 1.8558 & 1.8170 & $+0.0388$ & $+0.01$  & 0    &  4.0 & black hole \\
3 & 0.02   & 0.72405 & Type I   & 1.8308 & 1.7826 & $+0.0482$ & $-0.01$  & 0    &  4.0 & explosion to infinity \\
3 & 0.02   & 0.72405 & Type II  & 1.8432 & 1.7997 & $+0.0435$ & $+0.005$ & $-1$ &  4.0 & explosion to infinity \\
3 & 0.02   & 0.72405 & Type II  & 1.8432 & 1.7995 & $+0.0437$ & $+0.01$  & $-1$ &      & black hole \\
3 & 0.02   & 0.72405 & Type II  & 1.8435 & 1.7998 & $+0.0437$ & $-0.01$  & $-1$ &      & black hole \\
3 & 0.02   & 0.72405 & Type III & 1.8462 & 1.8017 & $+0.0445$ & $+0.01$  & $+1$ & 30.0 & black hole \\
3 & 0.02   & 0.72405 & Type III & 1.8462 & 1.7978 & $+0.0484$ & $+0.01$  & $-1$ & 30.0 & black hole \\
\hline
\end{tabular}
\caption{Results of simulations for $\ell=3$. Notice that in this case
  the maximum mass is $M_{\rm max} \simeq 2.245$.}
\label{table:l=3}
\end{table*}

\begin{table*}
\begin{tabular}{|c|c|c|c|c|c|c|c|c|c|c|}
\hline
$\ell$ & $a_0$ & $\omega$& Perturbation & $M$ & $N_B$ & $U$ & 
$\epsilon/ \varphi^{\rm max}_R $ & $s$ & $r_0$ & End result \\
\hline \hline
4 & 0.00005 & 0.84185 & Type 0   & 2.7458 & 2.8536 & $-0.1078$ & --       & --   &  --  & stable \\
4 & 0.00005 & 0.84185 & Type I   & 2.7531 & 2.8622 & $-0.1091$ & $+0.01$  & 0    & 11.5 & stable \\
4 & 0.00005 & 0.84185 & Type I   & 2.7385 & 2.8449 & $-0.1064$ & $-0.01$  & 0    & 11.5 & stable \\
4 & 0.00005 & 0.84185 & Type II  & 2.7459 & 2.8535 & $-0.1076$ & $+0.01$  & $-1$ & 11.5 & stable \\
4 & 0.00005 & 0.84185 & Type II  & 2.7459 & 2.8535 & $-0.1076$ & $-0.01$  & $-1$ & 11.5 & stable \\
4 & 0.00005 & 0.84185 & Type III & 2.7463 & 2.8540 & $-0.1077$ & $+0.01$  & $+1$ & 30.0 & stable \\
4 & 0.00005 & 0.84185 & Type III & 2.7462 & 2.8533 & $-0.1071$ & $+0.01$  & $-1$ & 30.0 & stable \\
4 & 0.0005  & 0.75793 & Type 0   & 2.6419 & 2.7181 & $-0.0762$ & --       & --   &  --  & black hole \\
4 & 0.0005  & 0.75793 & Type I   & 2.6539 & 2.7339 & $-0.0800$ & $+0.01$  & 0    &  7.5 & black hole \\
4 & 0.0005  & 0.75793 & Type I   & 2.6299 & 2.7024 & $-0.0725$ & $-0.01$  & 0    &  7.5 & migration to stable branch \\
4 & 0.0005  & 0.75793 & Type II  & 2.6419 & 2.7181 & $-0.0762$ & $+0.005$ & $-1$ &  7.5 & migration to stable branch \\
4 & 0.0005  & 0.75793 & Type II  & 2.6419 & 2.7180 & $-0.0761$ & $+0.01$  & $-1$ &  7.5 & black hole \\
4 & 0.0005  & 0.75793 & Type II  & 2.6420 & 2.7181 & $-0.0761$ & $-0.01$  & $-1$ &  7.5 & black hole \\
4 & 0.0005  & 0.75793 & Type III & 2.6430 & 2.7190 & $-0.0760$ & $+0.01$  & $+1$ & 30.0 & black hole \\
4 & 0.0005  & 0.75793 & Type III & 2.6430 & 2.7173 & $-0.0743$ & $+0.01$  & $-1$ & 30.0 & black hole \\
4 & 0.002   & 0.72290 & Type 0   & 2.3411 & 2.3117 & $+0.0294$ & --       & --   &  --  & black hole \\
4 & 0.002   & 0.72290 & Type I   & 2.3556 & 2.3318 & $+0.0238$ & $+0.01$  & 0    &  5.5 & black hole \\
4 & 0.002   & 0.72290 & Type I   & 2.3267 & 2.2918 & $+0.0349$ & $-0.01$  & 0    &  5.5 & explosion to infinity \\
4 & 0.002   & 0.72290 & Type II  & 2.3411 & 2.3116 & $+0.0295$ & $+0.005$ & $-1$ &  5.5 & explosion to infinity \\
4 & 0.002   & 0.72290 & Type II  & 2.3411 & 2.3115 & $+0.0296$ & $+0.01$  & $-1$ &  5.5 & black hole \\
4 & 0.002   & 0.72290 & Type II  & 2.3413 & 2.3118 & $+0.0295$ & $-0.01$  & $-1$ &  5.5 & black hole \\
4 & 0.002   & 0.72290 & Type III & 2.3430 & 2.3131 & $+0.0299$ & $+0.01$  & $+1$ & 30.0 & black hole \\
4 & 0.002   & 0.72290 & Type III & 2.3430 & 2.3104 & $+0.0326$ & $+0.01$  & $-1$ & 30.0 & black hole \\
\hline
\end{tabular}
\caption{Results of simulations for $\ell=4$. Notice that in this case
  the maximum mass is $M_{\rm max} \simeq 2.771$.}
\label{table:l=4}
\end{table*}


\subsection{Examples of our simulations}
\label{sec:num_examples}

In the Section we present some representative examples of our
numerical simulations. All the simulations shown here correspond to
the case of $\ell=2$.  For other values of $\ell$ the results are
qualitatively similar.

\begin{figure}
\includegraphics[width=0.49\textwidth]{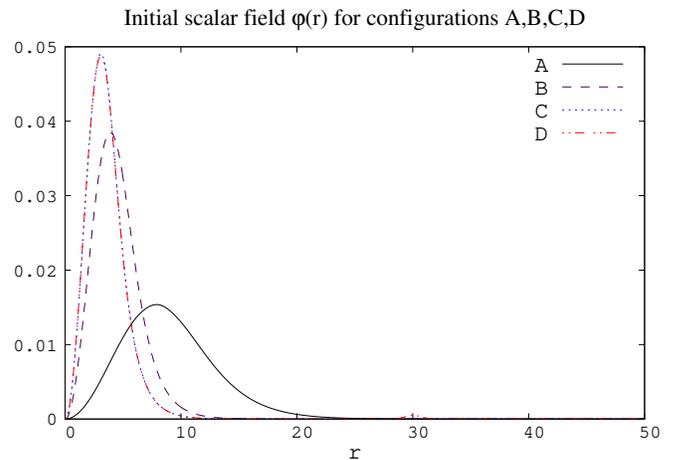}  
\caption{Initial value of the real part of the scalar field
  $\varphi_R(r)$ for the configurations $(A,B,C,D)$ of
  table~\ref{table:l=2}.  Notice that configurations $C$ and $D$ are
  almost identical, since they represent small (but different)
  perturbations of the same stationary solution.}
\label{fig:l=2-phi}
\end{figure}

\begin{figure}
\includegraphics[width=0.53\textwidth]{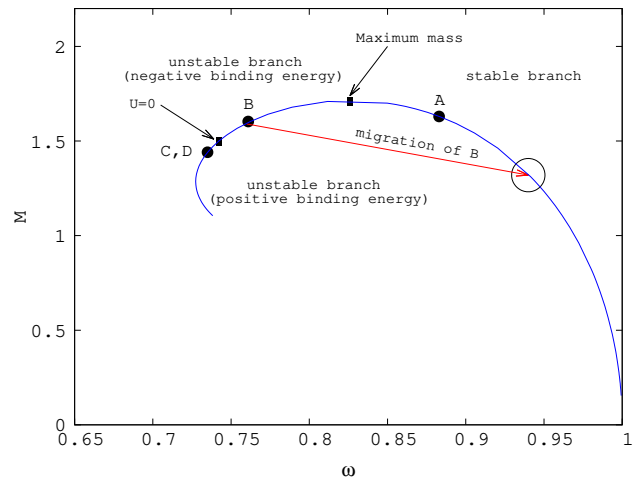}  
\caption{Initial positions on the $\ell=2$ mass-frequency diagram for the
  configurations $(A,B,C,D)$ of table~\ref{table:l=2}.  The black dots
  correspond to the initial (unperturbed) states, and the black
  squares mark the position of the maximum mass (minimum binding
  energy), and the place where the binding energy $U$ changes
  sign. Notice again that configurations $C$ and $D$ start from the
  same location, as they are different perturbations of the same
  stationary solution. Also shown is the approximate final state of
  configuration $B$ after it migrates to the stable branch (see text
  for details).}
\label{fig:migrate}
\end{figure}

We will show the results of four particular simulations, corresponding
to those configurations marked as $(A,B,C,D)$ in
table~\ref{table:l=2}. Figure~\ref{fig:l=2-phi} shows the initial
value of $\varphi_R(r)$ for these four configurations. Configuration
$A$ corresponds to a perturbation of a stable solution, configuration
$B$ to a perturbation of an unstable but bound solution, while
configurations $C$ and $D$ correspond to different perturbations of
the same unstable and unbound solution.  Notice that configurations
$C$ and $D$ are almost identical since they are different
perturbations of the same stationary solution.  Configuration $C$ adds
a small Gaussian close to the peak, while configuration $D$ adds one
outside the star at $r=30$ (this is barely visible in the plot).
Figure~\ref{fig:migrate} shows the initial position of these four
configurations on the $\ell=2$ mass-frequency diagram (compare with
figure~\ref{fig:summary} for $\ell=2$). In the figure we also show the
position of the maximum mass (minimum binding energy), and the place
where the binding energy $U$ changes sign. Also shown is the
approximate final state of configuration $B$ after it migrates to the
stable branch (see below).

\begin{figure*}
\includegraphics[width=0.99\textwidth]{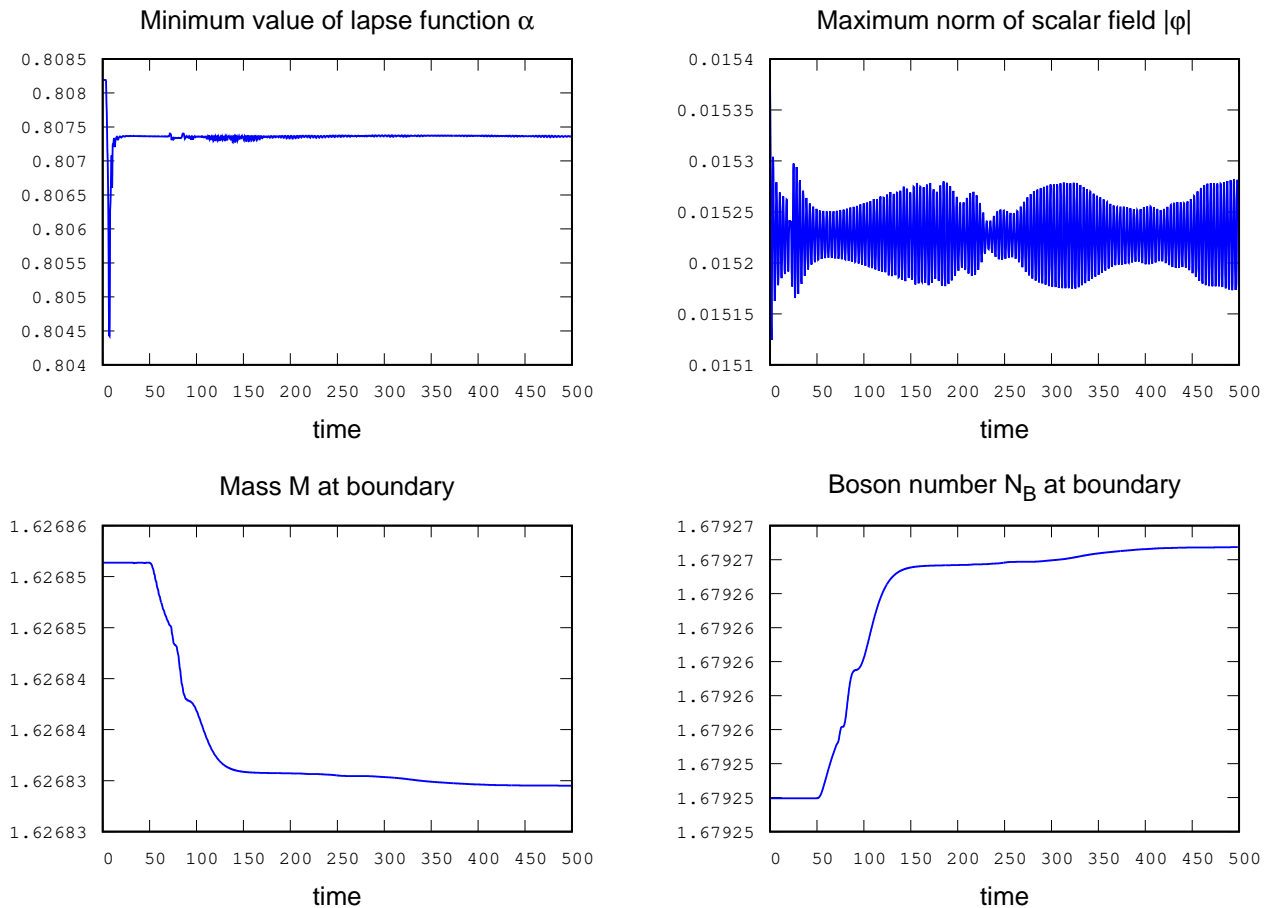}  
\caption{Evolution in time of configuration $A$. This is a stable
  configuration. The different panels show: Minimum of the lapse
  function $\alpha_{\rm min}$ (top-left), maximum norm of the scalar
  field $|\varphi|_{\rm max}$ (top-right), integrated mass $M$ at the
  outer boundary (bottom-left), and integrated boson number $N_B$ at
  the outer boundary (bottom-right).}
\label{fig:confA}
\end{figure*}

Let us now focus on configuration $A$, which corresponds to a
perturbation of type~II of a solution in the stable branch, with a
positive perturbation amplitude of $1\%$ at the peak of the scalar
field.  This configuration was run for $50,000$ times steps of size
$\Delta t=0.01$, resulting in a final time $T=500$. Some results for
this simulation are shown in figure~\ref{fig:confA}. The top-left
panel of the figure shows the minimum value of the lapse. We can see
that after an initial perturbation, it settles back down to a value
very close to the original one, and has very small oscillations for
the rest of the run. The top-right panel shows the value of the
maximum value of the norm of the scalar field $|\varphi| := \sqrt{\varphi
\varphi^*}$. Again we see that there are small oscillations around its
initial value.  Notice that for the stationary solution the norm is in
fact independent of time even if the scalar field is oscillating. The
bottom-left panel shows the value of the total integrated mass $M$ at
the boundary. Notice that initially it remains constant until $t \sim
50$. This is to be expected since for this run the boundary is located
precisely at $r=50$, and the scalar perturbation takes this long to
reach it. After this time, the mass decreases slightly and then
settles down to a smaller value. This indicates that a small pulse of
scalar field has been ejected by the star to infinity. Finally, the
bottom-right panel shows the total integrated boson number $N_B$ at
the boundary.  Again, we see that it remains constant until the
ejected pulse reaches the boundary at $t \sim 50$, it then increases
slightly and settles down to a higher value. This shows that the
ejected scalar field in fact has {\em negative} bosonic charge (in a
quantum mechanical interpretation it would be made of anti-particles).
The configuration is clearly stable, and after the initial
perturbation settles down to a new configuration very close to the
original one.

\begin{figure*}
\includegraphics[width=0.99\textwidth]{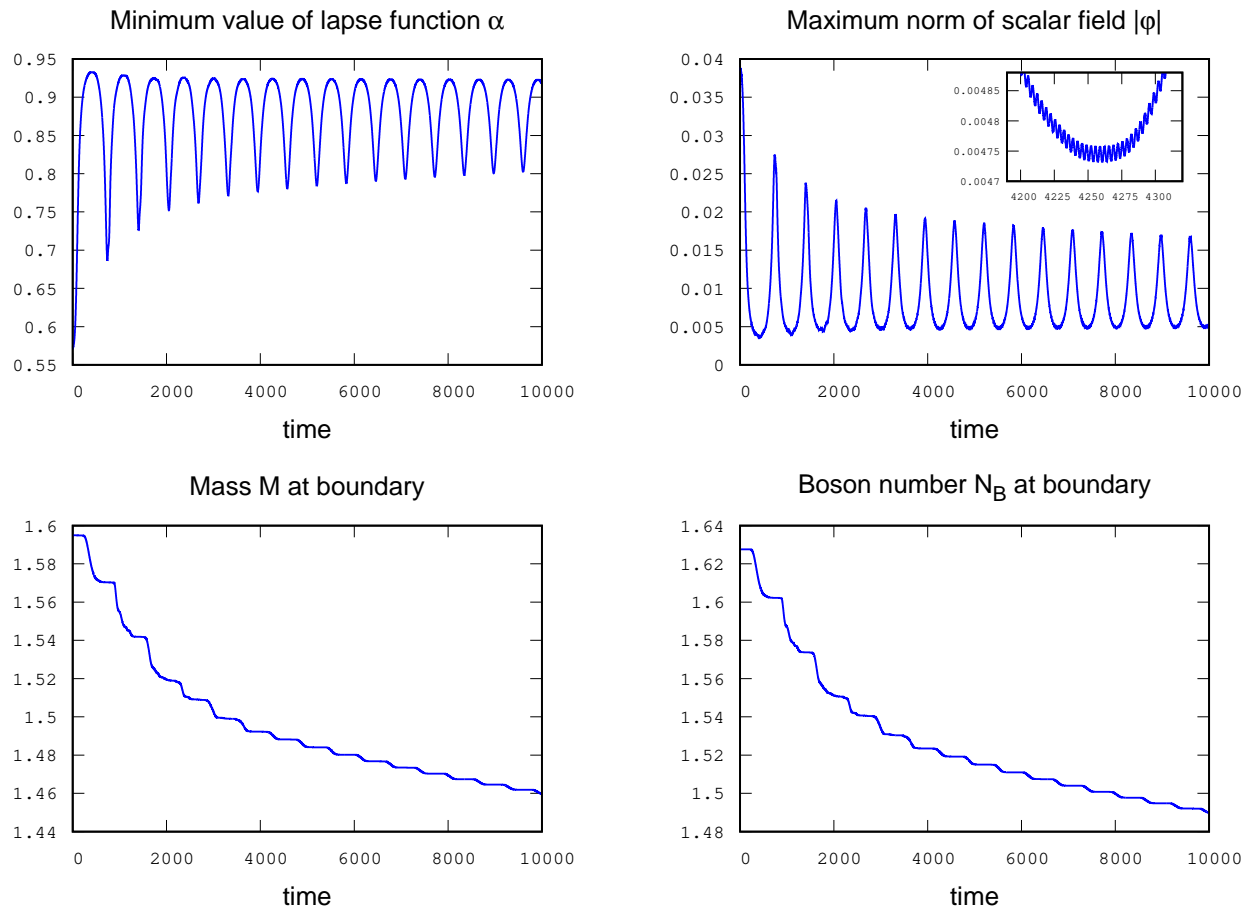}  
\caption{Evolution in time of configuration $B$. This configuration
  migrates to the stable branch. The different panels show: Minimum of
  the lapse function $\alpha_{\rm min}$ (top-left), maximum norm of
  the scalar field $|\varphi|_{\rm max}$ (top-right), integrated mass
  $M$ at the outer boundary (bottom-left), and integrated boson number
  $N_B$ at the outer boundary (bottom-right).  The inset to the
  top-right panel is a zoom to a small region of this plot to show the
  superposed small oscillations.}
\label{fig:confB}
\end{figure*}

Consider next configuration $B$, which corresponds to a type I
perturbation of an unstable but gravitationally bound solution. The
perturbation again has an amplitude of $1\%$ at the peak of the scalar
field, but in this case it is {\em negative}, that is, it decreases
slightly the size of the peak. This is an example of an unstable
solution that migrates to the stable branch. For this reason we have
in fact continued the simulation for a total of one million time
steps, reaching a final time of $T=10,000$. Results for this
simulation are shown in figure~\ref{fig:confB}, where the four panels
show the same quantities as before. The figure shows that the
evolution is now considerably more interesting.  Notice first the
minimum value of the lapse (top-left panel). It starts at a value of
$\sim0.57$, but rapidly increases and starts oscillating between $0.8$
and $0.93$.  These oscillations have a very long period of about
$\Delta T \sim 630$, corresponding to a frequency much smaller than
that of oscillations of the scalar field. The oscillations also seem
to be very slowly decreasing in amplitude, indicating that the
evolution will eventually settle down to a stationary configuration
after an extremely long time, though as mentioned before, at this
point we can not rule out the possibility that the configuration will
instead settle to some type of multi-oscillating
solution~\cite{Choptuik:2019zji}. Something very similar happens to
the maximum norm of the scalar field (top-right panel), which starts
at a value of $\sim 0.038$, and rapidly drops and starts oscillating
around $\sim 0.01$, with the same long period as the lapse. Here we
can also see some very small oscillations superposed to the large
ones, with a very short period, corresponding to the natural
oscillations of the scalar field (see inset in top-right panel for a
zoom of a small region of the plot). Again, the large oscillations
appear to be slowly decaying in amplitude.  When we look at the total
integrated mass $M$ and boson number $N_B$ (bottom two panels), we
notice that they are both decreasing in time, but they do so in steps
that become smaller and smaller with time.  They also seem to be
slowly converging to smaller values. The steps indicate that the boson
star is ejecting pulses of scalar field (with positive bosonic charge)
one at a time, with a period that matches the oscillations of the
lapse function.  The configurations clearly seems to have migrated to
the stable branch after ejecting excess scalar field in a series of
pulses, and is very slowly settling down.

For this particular configuration we in fact have also performed a
much longer simulation with 10 million time steps (reaching a final
time of $T=100,000$) in an attempt to determine the final state, but
we have found that at the end of this extremely long simulation the
configuration has yet to settle completely down. Our best estimate for
the final state is then only approximate.  By the end of this
simulation the total integrated mass is $M \sim 1.33$ and still slowly
falling, while the central value of the lapse is oscillating between
$0.87$ and $0.92$. Our best estimate of the final state can then be
obtained by assuming a final central lapse of $\alpha(r=0)\sim 0.9$,
which corresponds to a stationary solution with frequency $\omega \sim
0.94$, ADM mass $M \sim 1.31$ and total boson number $N_B \sim 1.33$
(this final configuration is shown as the open circle in
figure~\ref{fig:migrate}).

\begin{figure*}
\includegraphics[width=0.99\textwidth]{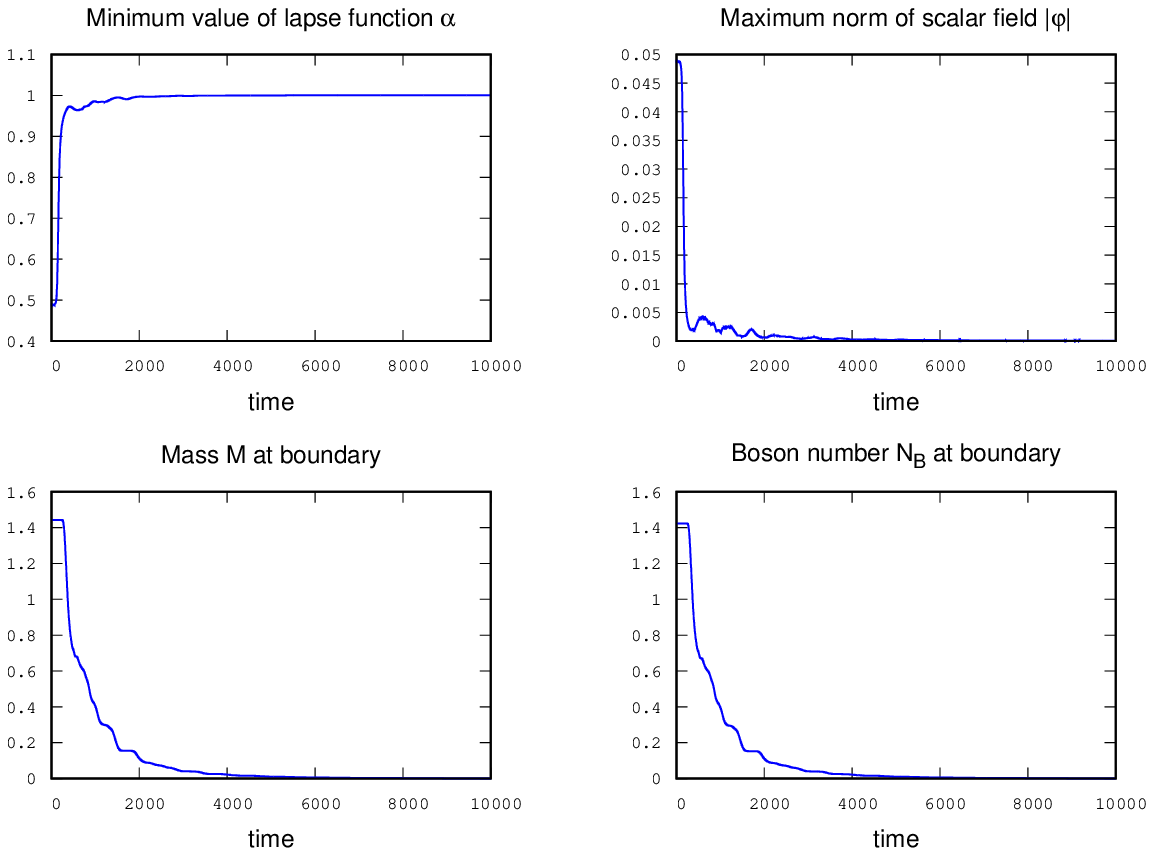}  
\caption{Evolution in time of configuration $C$. This configuration
  explodes to infinity. The different panels show: Minimum of the
  lapse function $\alpha_{\rm min}$ (top-left), maximum norm of the
  scalar field $|\varphi|_{\rm max}$ (top-right), integrated mass $M$
  at the outer boundary (bottom-left), and integrated boson number
  $N_B$ at the outer boundary (bottom-right).}
\label{fig:confC}
\end{figure*}

Let us now move to configuration $C$, which corresponds to a type II
perturbation of an unstable and unbound solution.  In this case the
perturbation adds a Gaussian with a small amplitude equivalent to only
$0.5\%$ of the maximum value of the scalar field. This is an example
of an unstable and unbound solution that explodes to infinity. Just as
we did for configuration $B$, we have again continued the simulation
for one million time steps, reaching a final time of
$T=10,000$. Results for this simulation are shown in
figure~\ref{fig:confC}.  Notice that the evolution is now very
different from that of configuration $B$.  The minimum of the lapse
grows rapidly from an initial value of $\sim 0.5$, and after a few
small oscillations becomes $1$, indicating that the spacetime is
essentially Minkowski. At the same time, the maximum norm of the
scalar field drops from its initial value, and after a few
oscillations goes to zero.  The total mass and boson number measured
at the boundary first remain constant until $T \sim 250$. They both
then drop rapidly, and after a series of steps also reach zero. The
scalar field corresponding to the boson star has then escaped
completely to infinity, leaving behind empty Minkowski spacetime.  The
fact that the total mass and boson number at the boundary only begins
to fall at $T \sim 250$ shows that there is a delay, and the boson
star does not begin to dissipate immediately, as otherwise one would
see effects at the boundary after one light-crossing time, that is $T
\sim 50$ (remember that the boundary is located at $r=50$).

Finally, consider configuration $D$.  This corresponds to a type III
perturbation of the {\em same} unstable and unbound solution of
configuration $C$. We now perturb the star with a small Gaussian with
an amplitude of $1\%$ of the maximum of the scalar field, but located
outside the star at $r=30$. This is now an example of an unstable and
unbound solution that collapses to a black hole.

\begin{figure*}
\includegraphics[width=0.99\textwidth]{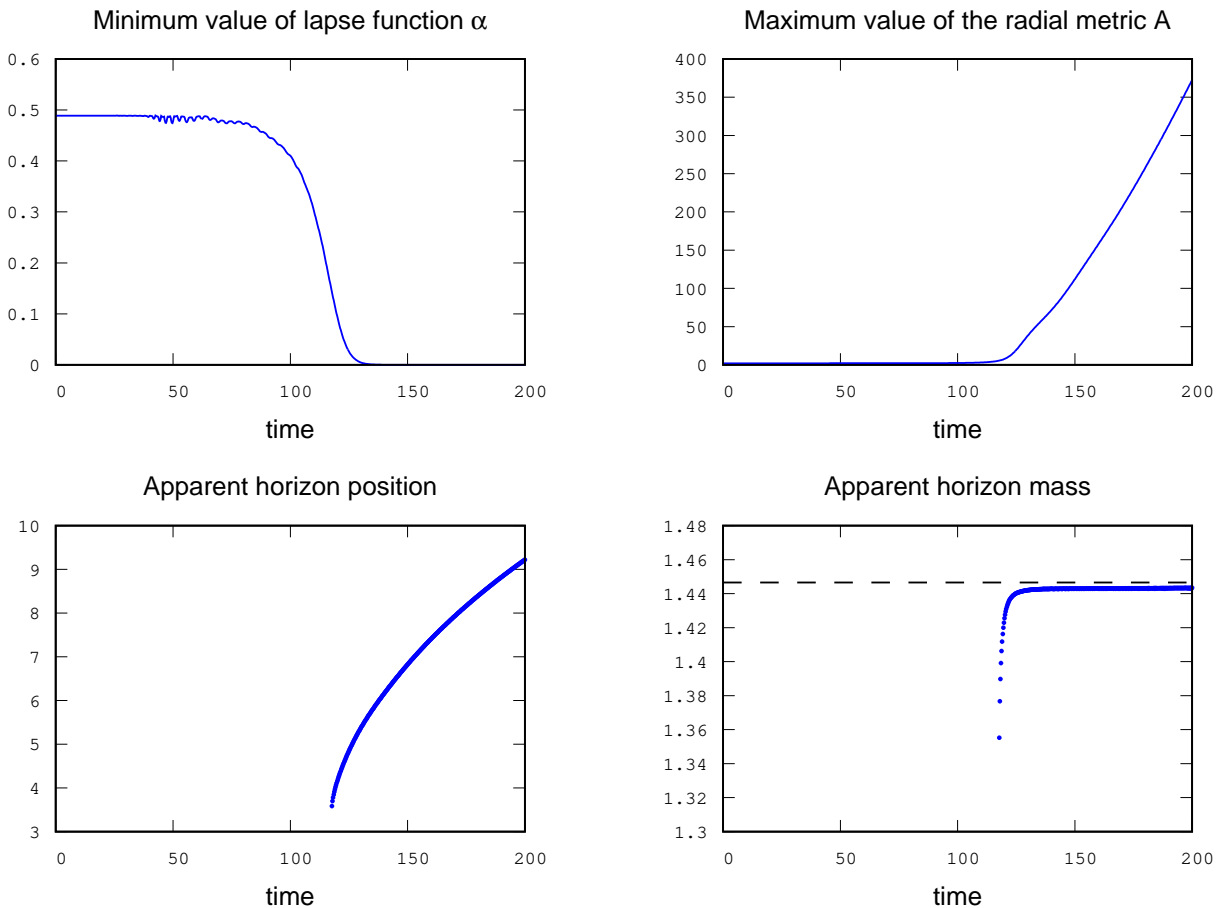}  
\caption{Evolution in time of configuration $D$. This configuration
  collapses to a black hole. The different panels show: Minimum of the
  lapse function $\alpha_{\rm min}$ (top-left), maximum value of the
  radial metric $A_{\rm max}$ (top-right), apparent horizon position
  (bottom-left), and apparent horizon mass (bottom-right).}
\label{fig:confD}
\end{figure*}

Now, the 1+log slicing condition that we use has the property of
``singularity avoidance'', that is, the lapse collapses to zero when a
black hole forms. Also, since we evolve with no shift, the collapse of
the lapse is accompanied by the well-known phenomenon of ``slice
stretching'', that is, the radial metric component grows rapidly close
to the black hole horizon (see for instance~\cite{Alcubierre08a}). 
All this implies that the integrated mass
and boson number accumulate large errors and stop being useful
quantities once the black hole forms (we are also approaching a
singularity, which makes matters worse). Because of this, we have
changed the quantities that we plot. We also only show the evolution
up to a final time of $T=200$, since after that the error associated
with the slice stretching effect start to become very large.  The four
panels of Figure~\ref{fig:confD} show the minimum value of the lapse
$\alpha$ in the top-left panel, the maximum value of the radial metric
$A$ (see equation~\eqref{eq:metric}) in the top-right panel, the
apparent horizon position in the bottom-left panel, and the apparent
horizon mass in the bottom-right panel.

Looking at the evolution of the minimum of the lapse we see that it
first remains constant for some time, until the initial perturbation
reaches the origin.  It then shows some small oscillations, and
finally, at $t \sim 100$, it starts to collapse rapidly to zero.  This
is an indication that a black hole has formed. The evolution of the
maximum value of the radial metric $A$ shows that it remains small and
constant until $t \sim 100$, and it then starts to grow rapidly
showing the typical behaviour of slice stretching. This is also
indicative of the formation of a black hole.

In order to make sure that a black hole has formed, we look for the
presence of an apparent horizon every 25 time steps. The bottom left
panel of figure~\ref{fig:confD} shows that an apparent horizon is
first found at $t \sim 117$.  Its initial coordinate radius is $r \sim
3.6$, but it then grows.  This growth is mostly just a coordinate
effect, as the physical horizon area rapidly becomes constant. This
can be seen in the bottom-right panel of the figure which shows the
apparent horizon mass (which is essentially the square root of the
area, see equation~\eqref{eq:BHmass}). The figure shows how once a
horizon forms, its mass first grows rapidly, and it then settles down to a
constant value of $M_H \sim 1.443$.  This is below the initial ADM mass of
the configuration, which in this case is $M \sim 1.447$ (shown as a
dashed horizontal line in the figure), indicating that a small amount
of scalar field has been lost to infinity.

\section{Conclusions}
\label{sec:conclusions}

We have performed a detailed study of the dynamical stability of the
recently proposed objects dubbed $\ell$-boson
stars~\cite{Alcubierre:2018ahf}. Through fully non-linear numerical
simulations we have shown that, just as it happens with the $\ell=0$
standard boson stars, for each value of $\ell$ the configuration of
maximum mass (which seems to coincide with that of minimum binding
energy) separates the parameter space into stable and unstable
regions. Stable configurations react to small perturbations by
oscillating and settling down to a new configuration close to the
original one, though this settling down process can be extremely
slow. Unstable configurations, on the other hand, can have three quite
different fates depending both on the specific type of perturbation
and on the sign of the total binding energy. For most types of
perturbations of unstable stars, collapse to a black hole is the most
likely outcome, regardless of the sign of the binding energy.
However, there are some regions of parameter space where perturbations
can result either in migration to a stable configuration if the total
binding energy is negative, or dispersion to infinity for a positive
binding energy. As mentioned before, for both stable configurations
and unstable configurations that migrate to the stable branch, the
relaxation times are extremely long, and we can not rule out the
possibility that those configurations will settle to some form of
multi-oscillating solution such as those studied
in~\cite{Choptuik:2019zji}.

We introduced three types of perturbations: type I is an internal
perturbation to the star that changes both the total mass and boson
density, type II is also an internal perturbation that preserves the
boson density to linear order in small quantities, and type three is
an external perturbation (scalar field falling into the boson
star) that always increases the mass but can either increase or
decrease the total boson number.

For unstable stars, type III perturbations always result in collapse
to a black hole, which is perhaps not surprising as they always
increase the total mass.  Type I and II perturbations can result
either in collapse to a black hole, or in migration/dispersion
(depending on the sign of the binding energy). The difference between
collapse and migration/dispersion seems to be related to whether the
perturbation increased or decreased the total mass of the original
configuration: if the mass was increased the configuration collapses,
while if it was decreased it can migrate/disperse.  Again, this is
perhaps to be expected.  However, we should mention the fact that
perturbations that result in migration/dispersion for small
amplitudes, result instead in collapse to a black hole if their
amplitude is increased beyond a certain (still small) value, even if
the sign and form of the perturbation remains the same. At this point
we have not been able to find a simple physical criterion that
predicts this change in behaviour.

We want to stress the role played by the $\ell$ parameter in our
configurations: As the value of $\ell$ grows, one finds more massive
and compact stable objects. This fact is consistent with the intuitive
idea that centrifugal effects in a rotating body oppose the
gravitational pull, so that one can have more massive stable objects
when compared to the non-rotating case.

It is interesting that $\ell$-boson stars represent a whole new family
of possible stable astrophysical objects. This encourages
observational searches for compact astrophysical objects, with
particular attention to features that could distinguish them from a
black hole.

We close this article with two remarks. First, we would like to
mention that linear perturbation theory, when applied to $\ell$-boson
stars, should allow one to study analytically some of the stability
results that we have discussed in this work. Work in this direction is
in preparation and will be presented elsewhere. Second, we stress that
in this work the perturbations of the $\ell$-boson star configurations
have been restricted to spherical symmetry. An important problem that
needs to be addresses is the extension of the stability analysis to
non-spherical perturbations, either by full 3D nonlinear simulations
of the Einstein-Klein-Gordon equations, or based on a linearized
perturbation analysis. The study of such perturbations for
$\ell$-boson stars should be particularly interesting, since in this
case even linearized perturbations may in principle transfer energy
between the different $\ell$ modes of the scalar field.


\acknowledgments

This work was supported in part by CONACYT grants No. 259228 and
182445, by the CONACyT Network Project No. 294625 ``Agujeros Negros y
Ondas Gravitatorias", by CONACYT Fronteras Project 281, by DGAPA-UNAM
through grants IN110218 and IA101318, by SEP-23-005 through grant
181345, by a CIC grant to Universidad Michoacana de San 
Nicol\'as de Hidalgo and by the European Union's Horizon 2020 research and innovation 
(RISE) program H2020-MSCA-RISE-2017 Grant No. FunFiCO-777740. A.B. and A.D.-T. were partially supported by PRODEP.


\bibliographystyle{apsrev}
\bibliography{referencias}


\end{document}